\documentclass[twocolumn,amsmath,amssymb,longbibliography,,superscriptaddress,orl,aps]{revtex4}

\usepackage{graphicx}
\usepackage{dcolumn}
\usepackage{amssymb}
\usepackage{bm}
\usepackage{color}
\usepackage{enumitem}
\usepackage{mathtools}
\usepackage{comment}
\usepackage{subfigure}
\usepackage{float}

\usepackage[colorlinks=true, urlcolor=blue, anchorcolor=blue, citecolor=blue,filecolor=blue,linkcolor=blue,menucolor=blue,pagecolor=blue]{hyperref}

\begin{document}

\title{Non-extensive super-cluster states  in aggregation with fragmentation}

\author{Nikolai V. Brilliantov}
\affiliation{Skolkovo Institute of Science and Technology, Moscow, Russia} 
\affiliation{Department of Mathematics,
University of Leicester, Leicester LE1 7RH, United Kingdom}
\author{Wendy Otieno}
\affiliation{Department of Mathematics, University of Leicester, Leicester LE1 7RH, United Kingdom}
\author{P. L. Krapivsky}
\affiliation{Department of Physics, Boston University, Boston, MA 02215, USA}
\affiliation{Skolkovo Institute of Science and Technology, Moscow, Russia}

\def\beq{\begin{eqnarray}}
\def\eeq{\end{eqnarray}}
\def\be{\begin{equation}}
\def\ee{\end{equation}}
\def\eq{&=&}
\def\ct{\citphysicale}
\def\bm{\begin{math}}
\def\me{\end{math}}
\def\bi{\bibitem}
\def\vrr{\vec r}
\def\vr{{\bf r}}
\def\hf{\frac{1}{2}}
\def\al{\alpha}
\def\rrb{\right )}
\def\llb{\left(}
\def\la{\langle}
\def\ra{\rangle}
\def\lbr{\left [}
\def\rbr{\right ]}
\def\del{\partial}
\def\grad{\nabla}
\def\ul{\underline}
\def\etal{{\it et al.}}
\def\lra{\leftrightarrow}
\def\rar{\rightarrow}
\def\lb{\label}
\def\bel{\begin{equation} \label}
\def\q{\quad}
\def\qq{\qquad}
\def\bel{\begin{equation} \label}
\def\beel{\begin{eqnarray} \label}
\newcommand \nn {\nonumber}
\newcommand \bei {\begin{itemize}}
\newcommand \eei  {\end{itemize}}
\newcommand \ii    {\item}
\newcommand \nt   {\nonumber \\ }

\begin{abstract}
Systems evolving through aggregation and fragmentation may possess an intriguing super-cluster state (SCS). Clusters
constituting this state are mostly large, so the  SCS resembles a gelling state. The formation of the SCS is
controlled by fluctuations and in this aspect, it is similar to a critical state. The SCS is non-extensive as
the number of clusters varies sub-linearly with the system size. In the parameter space, the SCS separates equilibrium
and jamming (extensive) states. The conventional methods such as the van Kampen expansion fail to describe the
SCS. To characterize the SCS, we propose a scaling approach with a set of critical exponents. Our theoretical findings
are in good agreement with numerical results.

\end{abstract}

\maketitle

Aggregation processes \cite{Smo16,Smo17,Chandra43} are ubiquitous in nature, social life,  and technology
\cite{Flory,Krapivsky,Leyvraz2003,Dorogov}. For instance, they underlie self-assembly, where pre-existing elemental
entities bind together due to local interactions \cite{SelfAssem,SA08}. Aggregation processes take place at diverse
temporal and spatial scales ranging from molecular scales \cite{Igor14,Erik17,RW04} and macroscopic scales where they
influence clouds and rain \cite{Seinfeld,Klett,Falkovich2002,Srivastava1982} to astrophysical scales where e.g.
aggregation of cosmic dust grains drives planetesimal and planetary ring formation
\cite{esposito2006,Bodrova2009,Guettler2010,Johansen,PNAS,Mazza,Blum}. Technological objects like swarm-bots also
demonstrate aggregation and self-assembling \cite{swarmbot}. In social networks, the merging units may be internet
users, enterprises, etc. \cite{Dorogov,Socnet1,Socnet2}.

In addition processes, the merging occurs only by addition of elemental units. Symbolically (see Fig.~\ref{PhaseDiagr})
\bel{eq:Add}
 \mathbb{M} + \mathbb{I}_k \xrightarrow{A_k} \mathbb{I}_{k+1}\,.
\ee
Here $\mathbb{M}\equiv \mathbb{I}_1$ denotes an elementary entity, a monomer, $\mathbb{I}_{k}$ is a cluster comprised of $k$ units, and $A_{k}$ is the rate of the process. Addition processes underlie self-assembly \cite{RW04,SA08,Privman2009,Igor14,Erik17}, internet  and business
systems. In material science, the addition mechanism dominates when the mobility of monomers greatly exceeds the
mobility of larger clusters \cite{bk, Blackman91, BlackmanMarshall_JPA94}.  This happens in several surface processes
when adatoms (monomers) diffuse on a substrate \cite{bk,Blackman91,BlackmanMarshall_JPA94,
Evans_PRB,Wolf,MBE,Zinke_Allmang1999, Krapivsky_EJB1998, Krapivsky_PRB1999, Family2001}, synthesis of nano-crystals
\cite{Privman2010,Privman2013}, aggregation of point defects in solids \cite{Koiwa,JNM2011}, etc. The Becker-D\"{o}ring
equation \cite{Ball1986,KingWattis2002,Niethammer2003,Wattis2006,Wattis2009} and the
Lifshitz-Slyozov-Wagner model also rely \cite{Niethammer1999,Herrmann2009} on the addition mechanism.

Aggregation is often accompanied by cluster disintegration that may occur, e.g., due to the accumulation of faulty steps in self-assembling. Disintegration can proceed spontaneously \cite{Ball1986,KingWattis2002,Niethammer2003,Wattis2006,Wattis2009,Niethammer1999}  or be caused by interactions with monomers that trigger either addition or disintegration. The reaction scheme
\bel{eq:Break}
\mathbb{M}+\mathbb{I}_{k} \xrightarrow{S_{k,\boldsymbol{\ell}} }
\mathbb{M}+\underbrace{\mathbb{I}_{l_1}+\ldots+\mathbb{I}_{l_n}}_{|\boldsymbol{\ell}|=l_1+\ldots +l_n =k}
\ee
represents the breakage into the debris $\boldsymbol{\ell}=\{l_1, \ldots l_n\}$. The collision-controlled fragmentation underlies e.g. the Oort-Hulst models \cite{Laurencot2001,OortHulst,Laurencot2007,Wattis2012,Dubovski1999}. Generally, the process \eqref{eq:Break} describes the break-up of an aggregate in a collision with energetic monomers
\cite{PNAS,stadnichuk2015smoluchowski,MKSTB,KOB,ColmLast}. The complete breakage
\bel{eq:Shat} \mathbb{M}+\mathbb{I}_{k} \xrightarrow{S_k} \underbrace{\mathbb{M}+\ldots+\mathbb{M}
}_{k+1}\,.
\ee
is known as the shattering process \cite{Guettler2010,BlumErosion2011,PNAS,Krapivsky}; it is included in the Oort-Hulst
models.  Qualitatively similar behaviors emerge for partial \eqref{eq:Break} and complete \eqref{eq:Shat} breakage,
provided that a large number of elementary units is produced. Here we present the analysis for the shattering model
\eqref{eq:Shat}; the results for the general model \eqref{eq:Break} are given in the Supplementary Material (SM) \cite{SM}.

Here we investigate addition-shattering processes and observe rich behaviors. Besides the equilibrium states (ESs) and
jammed states (JSs), we reveal intriguing super-cluster states  (SCSs) composed of mostly very large clusters. The SCSs
are non-extensive --- the number of emerging structures does not scale linearly with the system size; furthermore,
fluctuations play a dominant role there. Conventional approaches fail to describe the SCS and we propose a framework to
characterize it. Below detailed definitions of JSs and SCSs are  given.
\begin{figure}
\centering
\includegraphics[width=4.15cm]{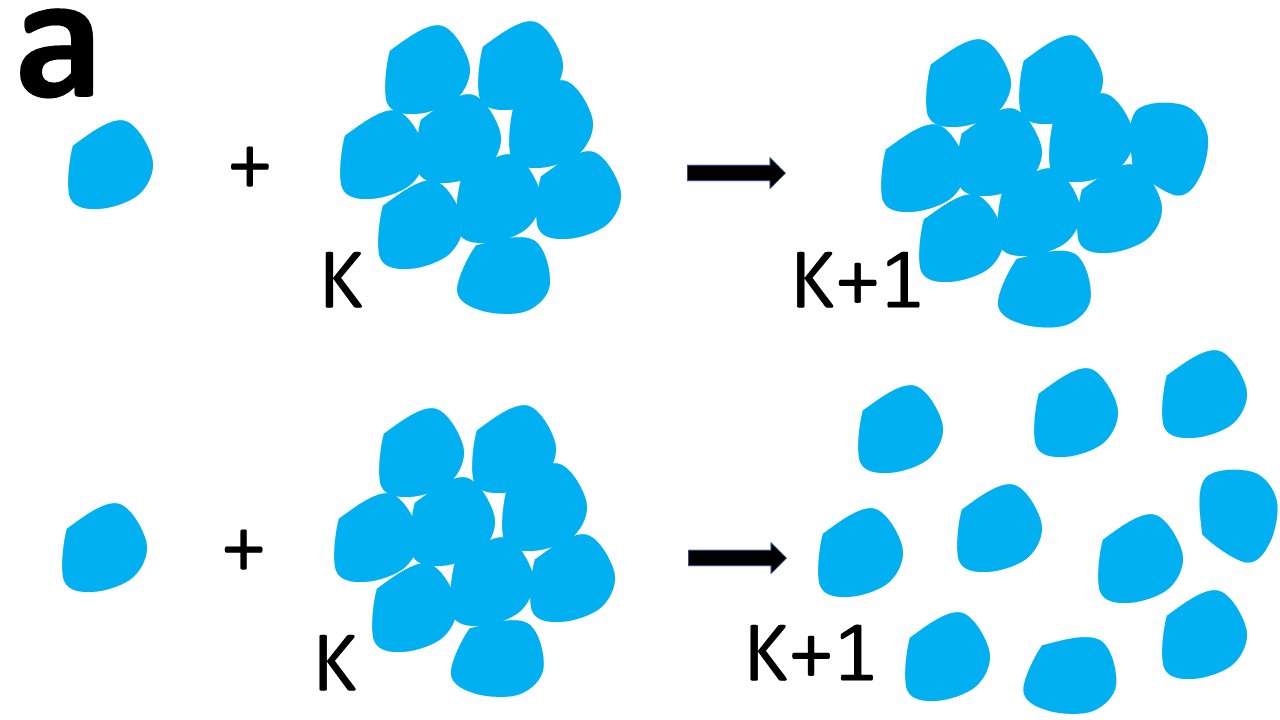} \,
\includegraphics[width=4.15cm]{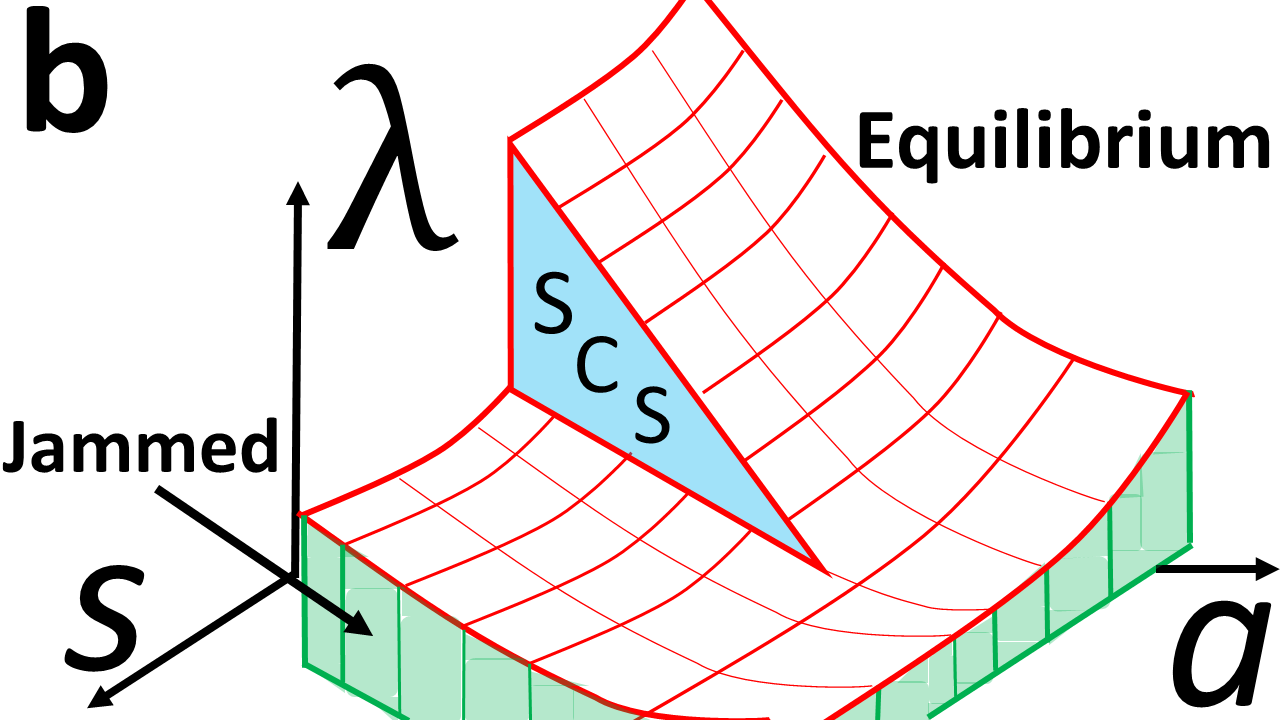}
\caption{(a) Additional aggregation with disintegration. (b) Schematic phase diagram of aggregating systems with
disintegration in $(a,s,\lambda)$ domain. The non-extensive SCS lie on a surface that is surrounded by the extensive
jammed states (JSs) and equilibrium steady states (ESSs). }
\label{PhaseDiagr}
\end{figure}

Addition and shattering rates often vary algebraically with the aggregate size. We thus consider the rates
\begin{equation}
\label{eq:ASdef}
A_k = k^a, \qquad  \qquad S_k = \lambda k^s.
\end{equation}
The amplitude in addition rate is set to unity by time re-scaling. The dependence \eqref{eq:ASdef} with $a \leq 1$
simply reflects the fact that the aggregation rate is proportional to the  clusters surface (which may be fractal); the
aggregation in networks also obeys \eqref{eq:ASdef} with $a \leq 1$ \cite{Dorogov,KrapRedner2001}. The intensity of the
shattering process is quantified by $\lambda$, while the exponent $s$ depends on its mechanism; commonly $s \leq 1$.

Denote by $c_k(t)$ the density of aggregates of size $k$. With rates \eqref{eq:ASdef},
the governing equations read
\begin{subequations}
\label{eq:Gen}
\begin{align}
 \label{c1tau}
\frac{dc_{1}}{d t} &= -c^2_{1} - \sum_{j=1}^{\infty} j^{a} c_1c_{j}+ \lambda \sum_{j=2}^{\infty} j \cdot j^{s} c_1c_{j} \,,\\
\frac{dc_{k}}{d t} &= c_1[(k-1)^{a}c_{k-1} - k^{a}c_{k}] - \lambda k^{s} c_1c_{k}\,.
\label{cktau}
\end{align}
\end{subequations}
Equation \eqref{cktau} is valid for all $k\geq 2$. The right-hand side of Eq.~\eqref{c1tau} reflects that the monomer density
decreases  due to aggregation with other monomers and clusters (first and second terms) and increases due to shattering.

First, we illustrate the generic behavior of the system on tractable models. Then a conjecture about its general
behavior is confirmed numerically.

\emph{Model with $(a,s)=(1,0)$.} In terms of the modified time, $\tau = \int_{0}^{t} c_{1}(t') dt'$,
Eqs.~\eqref{c1tau}--\eqref{cktau} linearize
\begin{subequations} \label{eq:s0a1}
\begin{align}
\dot{c}_1 &= -(1+\lambda)c_{1} - 1 + \lambda,
\label{a1s0c1tau} \\
\dot{c}_k&= (k-1)c_{k-1} - (k + \lambda)c_{k}, \quad k\geq 2.
 \label{a1s0cktau}
\end{align}
\end{subequations}
Hereinafter  $\dot{f} \equiv df/d\tau$.  We choose the units where  the mass density conservation reads
$M_1=\sum_{k\geq 1} k c_k=1$. Solving \eqref{a1s0c1tau} for the most natural mono-disperse initial conditions, $
c_k(0)=\delta_{k,1}$, we obtain
\begin{equation}
\label{a1s0c1tausol} c_{1}(\tau) = \frac{2}{1+\lambda} e^{-(1+\lambda)\tau} - \frac{1-\lambda}{1+\lambda}
\end{equation}
This exact result shows that different behaviors emerge depending on whether $\lambda$ is less than, equal to, or
larger than $\lambda_{c} = 1$:  If $\lambda \geq \lambda_c=1$, the monomer density $c_1(\tau)$ always remains
non-negative, while for $\lambda <1$, the monomer density formally becomes negative as a function of the modified time.
The requirement $c_1 \geq 0$ implies the existence of $\tau_\text{max}$ such that the system evolves only until $\tau
\leq \tau_\text{max}$, where $c_1(\tau_\text{max})=0$; the modified time $\tau_\text{max}$ corresponds to the infinite
physical time $t=\infty$ \cite{SM}.

In the subcritical case, $\lambda<1 $, the relation between $t$ and $\tau$ is found from \eqref{a1s0c1tausol}
yielding
\begin{equation*}
c_1(t)= \frac{1-\lambda}{2 e^{(1-\lambda)t} - 1-\lambda}\,.
\end{equation*}
Thus the monomer density vanishes at $t \to \infty$ if $\lambda<1$. Other cluster densities
remain positive. Near the critical point ($0<1-\lambda\ll 1$), they simplify to \cite{SM}
\begin{equation}
\label{Ck10:sub} c_k (\infty)= \left[\frac{1}{2}-\frac{1}{k(k+1)}\right](1-\lambda) + O[(1-\lambda)^{3/2}].
\end{equation}
Final densities depend on the initial condition, see  Fig.~\ref{Fig:C1NCK}.  Hence, for $\lambda<1$ the system falls
into a jammed state --- a non-equilibrium stationary state, with a structure depending on initial conditions
\cite{Jam1,Jam2}. There are no monomers vanish in the JSs.

At the critical point $c_k = e^{-2\tau}\left(1-e^{-\tau}\right)^{k-1}$ if $c_k(0)=\delta_{k,1}$.
From $c_1(\tau)=e^{-2 \tau}$, we get $2\tau= \ln(1+2t)$ and
\begin{equation}
\label{10:crit} 
c_k = \frac{1}{1+2t}\left[1-\frac{1}{\sqrt{1+2t}}\right]^{k-1}, \quad c = \frac{1}{\sqrt{1+2t}},
\end{equation}
where $c=\sum_{k\geq 1} c_k$ is the total cluster density. All densities vanish at $t=\infty$ independently on initial
conditions, yet the mass density is conserved, $M_1=1$. The same is true for pure aggregation, where a single cluster
(gel) is eventually formed in a finite-size system. As we show below, the ultimate state  for a finite size system
dramatically differs here: The final number of clusters varies from realization to realization and its average scales
sub-linearly with the system size. We call such states super-cluster states (SCSs), providing a precise definition
below. The SCSs manifest themselves by the vanishing densities $c_k(\infty)$ for all $k$ in the thermodynamic limit.

In the supercritical regime $\lambda>1$, the cluster densities relax exponentially fast to the equilibrium steady state
that does not depend on initial conditions \cite{SM}:
\begin{equation}
\label{10_final} c_k(\infty) = (\lambda-1)\,\frac{\Gamma(k)\,\Gamma(1+\lambda)}{\Gamma(k+1+\lambda)} \, .
\end{equation}
Equations \eqref{Ck10:sub}--\eqref{10_final} demonstrate that $c_k(\infty) \to 0$ when $\lambda \to 1\pm 0$, indicating that at $\lambda=1$
the system undergoes  a continuous phase transition from the jammed state to the equilibrium steady state through the critical SCS with vanishing densities, see Fig.~\ref{Fig:C1NCK}.

\begin{figure}
\centering
\includegraphics[width=4.25cm]{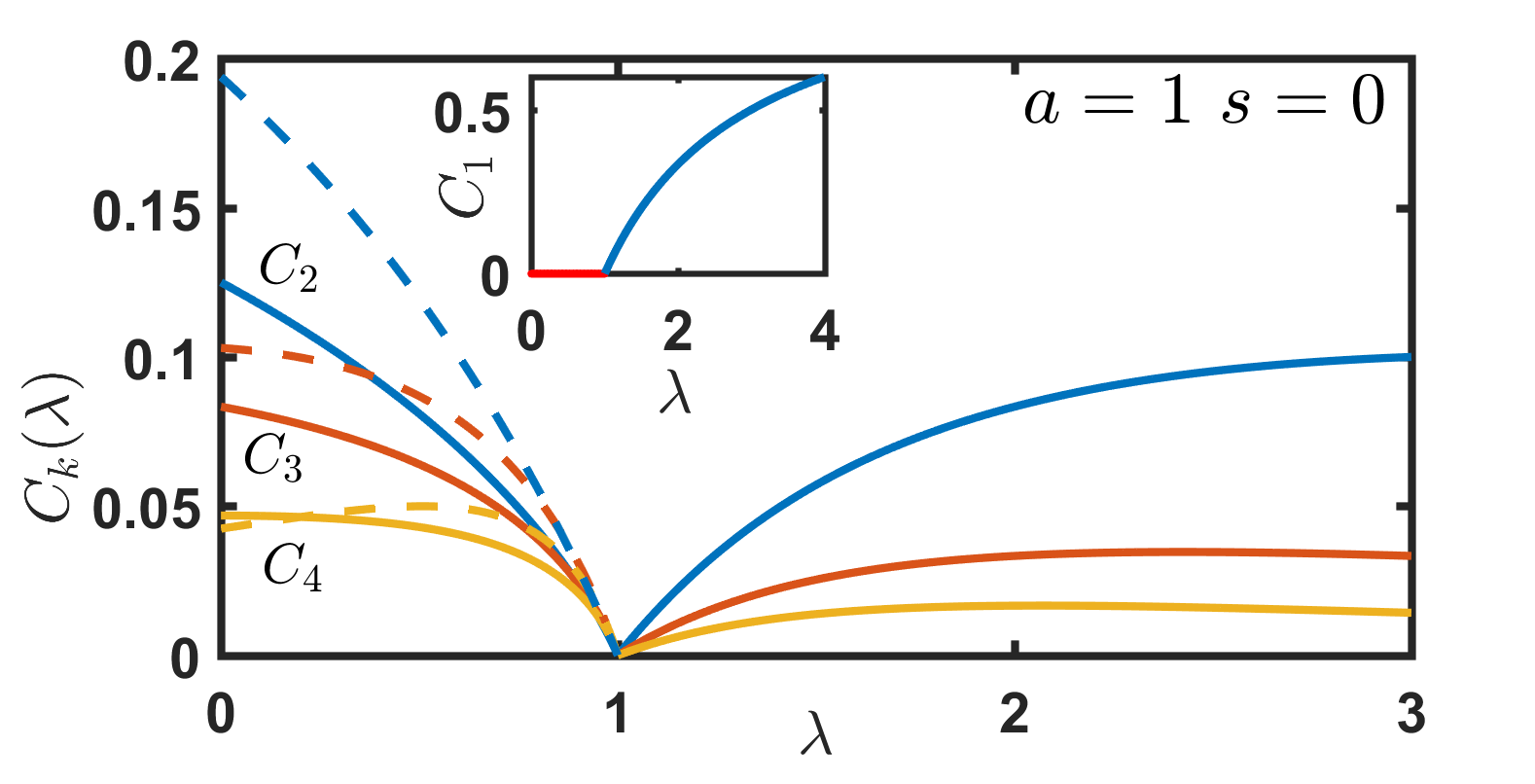}
\includegraphics[width=4.25cm]{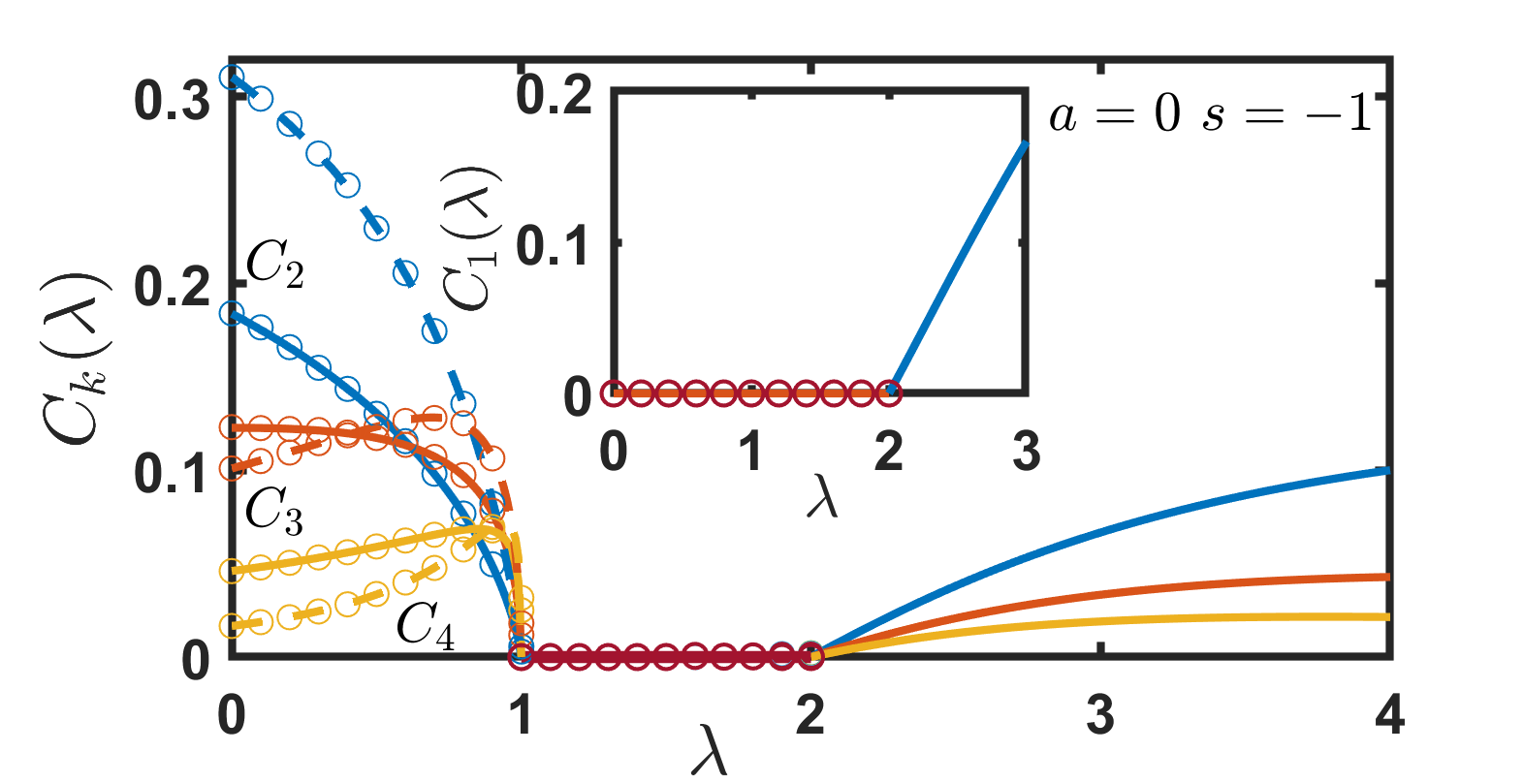}
\caption{Left panel: The final densities $c_k(\lambda)\equiv c_k(t=\infty)$ versus $\lambda$ for the model
$(a,s)=(1,0)$. Initial conditions are mono-disperse (solid lines); monomer-dimer, specifically $c_{1}(0) = 0.2$ and
$c_{2}(0) = 0.4$, (dashed lines). Right panel: The same for the model $(a,s)=(0,-1)$. Curves: analytical (for $\lambda
\geq 1$) and numerical (for $\lambda <1$) solutions of rate equations;  dots: Monte Carlo (MC) results. The system size
is $N=10^6$. All densities vanish in the SCS at $\lambda=1$ (left panel) and  $1 \leq \lambda \leq 2$ (right panel).
Insets: The final density of monomers $c_1(\lambda)$. }
 \label{Fig:C1NCK}
\end{figure}

\emph{Model with $(a,s)=(0,-1)$. } The rate equations read
\begin{subequations}
\label{eq:s0a-1}
\begin{align}
\dot{c}_1&= - (1+ \lambda)c_{1} + (\lambda - 1)c,\\
\dot{c}_k&= c_{k-1} - ( 1+ \lambda/k) c_{k} \quad k\geq 2, \label{ckpt1a0sm1}
\end{align}
\end{subequations}
The model with $\lambda_c=1$ again demarcates different evolution regimes. In the subcritical regime, $\lambda<1$, the
system falls into a jammed state with vanishing monomer density, $c_1( \tau_\text{max})=0$; the final cluster densities
$c_k(\tau_\text{max})$ are determined by initial conditions, see Fig.~\ref{Fig:C1NCK}.

At the critical point, $\lambda=1$, the solution for $t \gg 1$ reads \cite{SM}
\begin{equation}
\label{ck:0-1-tau} c_k(t)\simeq k^{k-1} (1+2t)^{-\frac{k+1}{2k}},
\end{equation}
indicating that all densities vanish at the critical point, see Fig.~\ref{Fig:C1NCK}. When $\lambda>1$, the Laplace
transforms of the densities is obtained iteratively from Eqs.~\eqref{eq:s0a-1} to give
\begin{equation}
\label{eq:ckpHG}
\widehat{c}_k(p)=\frac{1}{p}\,\frac{(1+\lambda\epsilon) \epsilon^k}{F\left[2,2;2+\lambda\epsilon;
\epsilon\right]}\, \frac{k!\,\Gamma(1+\lambda \epsilon)}{\Gamma(k+1+\lambda\epsilon)}
\end{equation}
where $\widehat{c}_k(p)=\int_0^{\infty} c_k(\tau)e^{-p\tau}d \tau$ and  $\epsilon=(1+p)^{-1}$. The hypergeometric
function appearing in \eqref{eq:ckpHG} admits an integral representation
\begin{equation}
\label{int} F\left[2,2;2+\lambda\epsilon; \epsilon\right] = \lambda\epsilon(1+\lambda\epsilon) \int_0^1
dx\,\frac{x(1-x)^{\lambda\epsilon-1}}{(1-x\epsilon)^2}.
\end{equation}
Using \eqref{eq:ckpHG}--\eqref{int}, one can extract the asymptotic behavior of $c_k(\tau)$ at $\tau \to \infty$,
from the behavior of $\widehat{c}_k(p)$ at $p \to 0$. For $\lambda >2$  the function $F$ is regular at $p=0$ and equals to
$\lambda(1+\lambda)/(\lambda-2)$. The Laplace transform $\widehat{c}_k(p)$ has a simple pole, $\widehat{c}_k(p) \to c_k(\infty)/p$ as $p \to
0$, indicating the existence of a steady state size distribution, $c_k(\infty)$.

Within the critical interval $1 \leq \lambda\leq 2 $ the function $F\left[2,2;2+\lambda\epsilon; \epsilon\right]$
diverges as $p \to 0$ implying $c_k \to 0$ for $\tau \to \infty$. Overall, the final densities read
\begin{equation}
\label{eq:Cka0sm1}
c_k(\infty) =
\begin{cases}
c_k(\tau_\text{max})(1-\delta_{k,1})  & \lambda < 1 \\
0                                            & 1 \leq \lambda\leq 2\\
\frac{k!\,(\lambda-1)(\lambda-2)\Gamma(\lambda)}{\Gamma(k+1+\lambda)} & \lambda > 2,
\end{cases}
\end{equation}
with $c_k(\tau_\text{max})$ depending on initial conditions. The system undergoes continuous phase  transitions from a
JS to a SCS at $\lambda=\lambda_{\rm low}=1$ and from a SCS to an ES at $\lambda=\lambda_{\rm up}=2$. The cluster
densities decay algebraically when $1<\lambda <2$ and logarithmically when $\lambda=2$.

\emph{Models with $s=a-1$.}
The rate equations read
\begin{subequations}
\label{eq:s2a}
\begin{align}
\label{1ASa}
\dot{c}_1&= - (1+\lambda)c_1 + (\lambda-1)M_a, \\
 \label{kASa}
\dot{c}_k&= (k-1)^a c_{k-1}- k^a(1+\lambda/k)c_k, \quad k\geq 2,
\end{align}
\end{subequations}
with $M_a=\sum_{k\geq 1} k^a c_k$. The SCS occurs \cite{SM} when
\begin{equation}
\label{lambda_down-up}
1=\lambda_\text{low} \leq \lambda \leq  \lambda_\text{up}=2-a
\end{equation}
and the final densities are
\begin{equation}
\label{eq:Ck_gen_a} c_k(\infty) =
\begin{cases}
(1-\delta_{k,1}) c_k(\tau_\text{max})  & \lambda < 1 \\
0                                                        & 1 \leq \lambda\leq 2-a\\
\frac{k^{-a} k!/\Gamma(k+\lambda+1) }{\sum_{n \geq1} n^{1-a}n! /\Gamma(n+\lambda+1)} & \lambda > 2-a.
\end{cases}
\end{equation}

Thus for the three-parameter class of models \eqref{eq:ASdef}, the SCS (characterized by $c_k(\infty)=0$) emerges when
$s = a-1$ and $1 \leq \lambda \leq 2-a$, with a continuous phase transition from the SCS to the JS at $\lambda =1$, and
to the ES at $\lambda =2 -a$. The relaxation to the JS and ES  is exponentially fast, while to the SCS is algebraic in
time, when $1\leq \lambda<2-a$, and logarithmic for  $\lambda_{\rm up}=2-a$, see \cite{SM}.

In the SM we show that the emergence of SCSs is robust to incomplete shattering, provided that monomers are abundantly
produced. For instance, it occurs if only half of a cluster disintegrates into monomers. The appearance of SCSs
requires a faster growth with the cluster size of the aggregation rate than of the fragmentation rate. The latter
however  should be large enough to provide abundant monomers feeding  the large clusters.

A detailed analysis shows that at $\lambda=\lambda_{\rm low}=1$, the system  undergoes an infinite sequence of weak
first-order phase transitions (see SM). They occur at critical values $a_1=1$, $a_2=0.415$, $a_3=0.224$, etc., and are
manifested by an abrupt change of the relaxation kinetics of the cluster densities \cite{KrapNew}, see \cite{SM}.

Monomers also play a key role in Becker-D\"{o}ring models with evaporation and Oort-Hulst models,  yet the production
of monomers never ceases in these models and hence the jammed and super-cluster states do not emerge.

\emph{The nature of the SCS.} To understand the difference between SCSs  and gelling states we consider large, but
finite systems of $N \gg 1$ monomers. Denote by $C_k(t)$ the total number of clusters of size $k$ and by $C(t)$ the
total number of clusters. The densities $c_k(t) = C_k(t)/N$ and $c(t)=C(t)/N$ usually do not depend on the system size
when $N\gg 1$. The rate Eqs. \eqref{eq:Gen} describe the evolution for $c_k(t)$, but they can fail, as the usage of the
densities is based on the tacit assumption that the behavior is extensive. Generally, finite stochastic systems
are explored by explicitly modeling each elementary reaction. That is, in a single reaction event a configuration
$(C_1,C_2,\ldots C_N)$ transforms into one of the following:
\begin{subequations}
\label{MC}
\begin{align}
\label{channel:m}
& (C_1-2,C_2+1)     ~\quad \qquad \qquad \text{rate} ~~C_1(C_1-1)/N,\\
\label{channel:add}
&(C_1-1,C_k-1,C_{k+1}+1)    \quad \text{rate} ~~k^aC_1 C_k/N,\\
\label{channel:shat} &(C_1+k,C_k-1)    ~\quad \qquad \qquad \text{rate} ~~k^sC_1 C_k/N.
\end{align}
\end{subequations}
(Only the components of an evolved configuration that differ from the original configuration are shown.) The reaction
rates correspond to the rates \eqref{eq:ASdef}  and accounts automatically  for the finiteness of the system. The
quantities $C_k(t)$ are random variables and the system is characterized by the averages $\langle C_k(t) \rangle$,
$\langle C_k(t) C_j(t)\rangle$, etc.

\begin{figure}
\centering
\includegraphics[width=4.3cm]{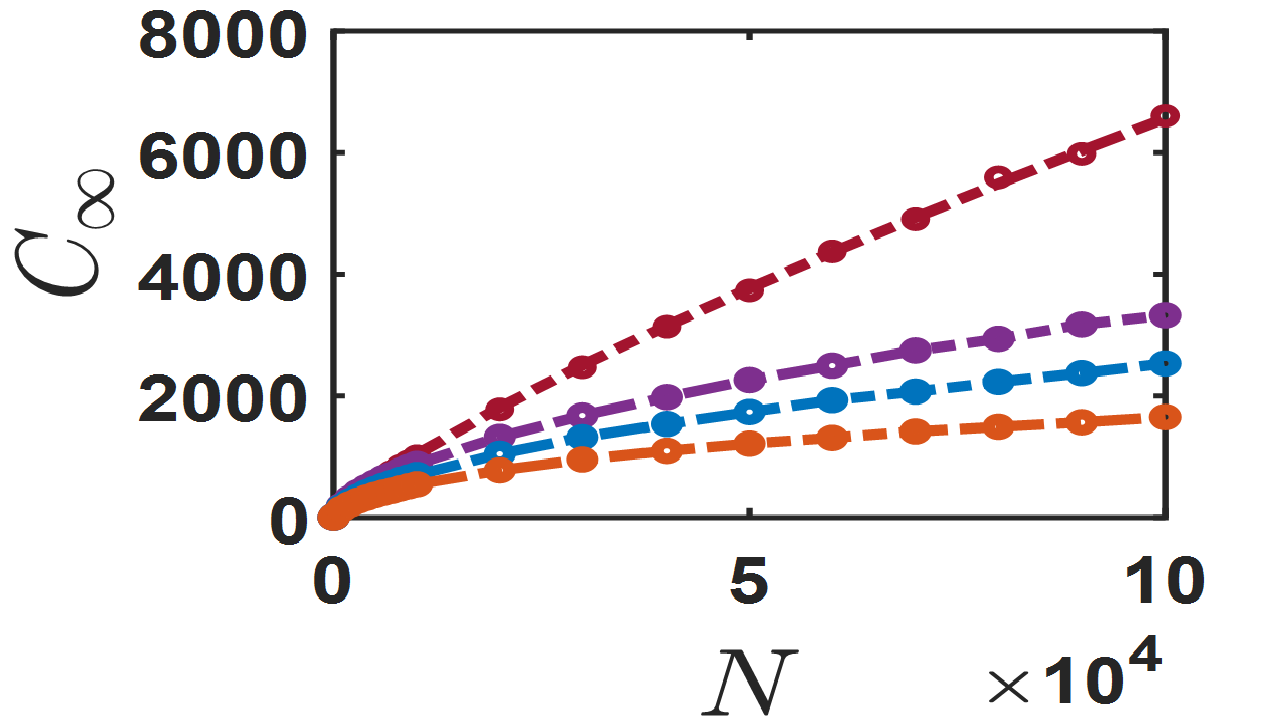}\,\,
\includegraphics[width=4.0cm]{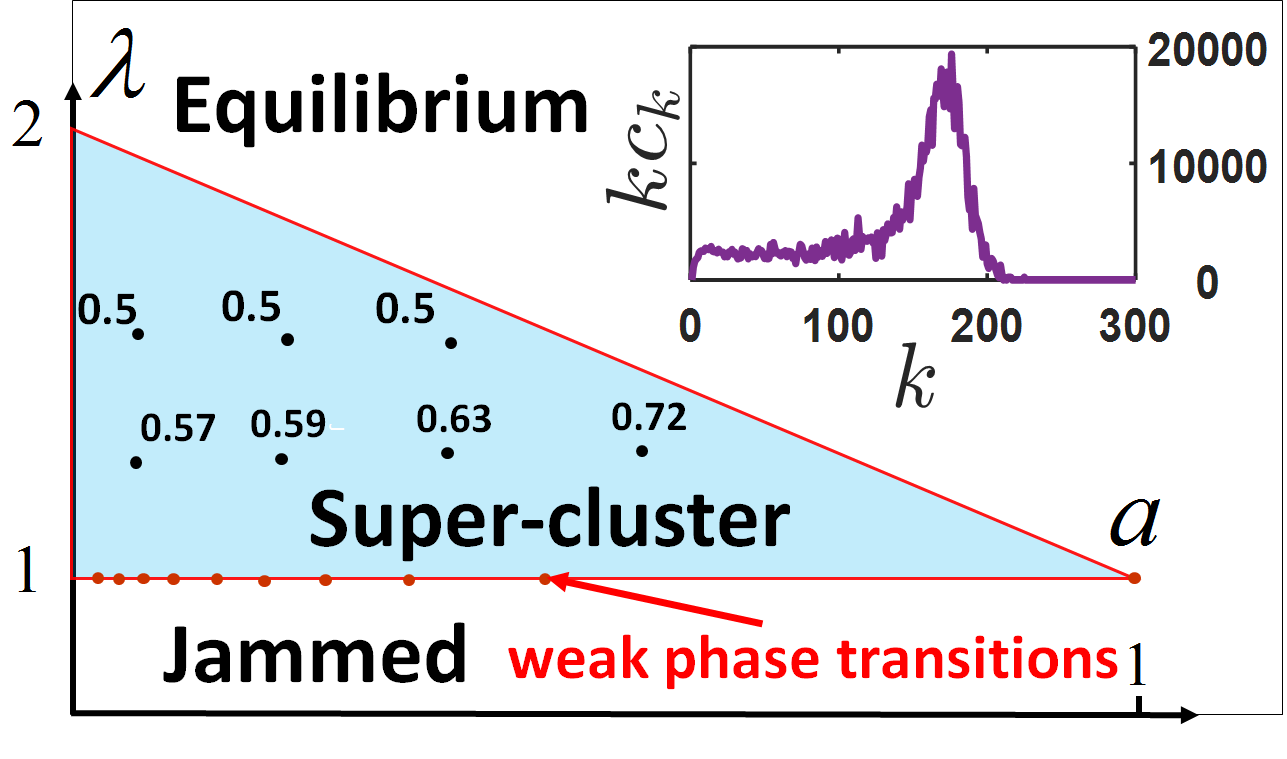}
\caption{Left panel: The total number of clusters in the final SCS versus $N$. MC results are shown by dots; fits for
the scaling law, $C(\infty) \sim N^{\delta}$, are shown by lines. Curves (top to bottom): $(a,s,\lambda) =(1,0,1)$ with
$\delta =4/5$,  see Eq.~\eqref{clusters-av:final}; $(a,s,\lambda) =(0,-1,125)$ with $\delta =0.571$; $(a,s,\lambda)
=(0,-1,135)$ with $\delta =0.599$; $(a,s,\lambda) =(0,-1,1.5)$ with $\delta =0.5 $. Right panel: SCS in the
$(a,\lambda)$ domain. It borders JSs at $\lambda_{\rm low}=1$ and ESSs at $\lambda_{\rm up} = 2-a$. The black dots with
numbers indicate the values of $\delta$. The red dots indicate the points of the weak first-order phase transitions.
Inset: The mass distribution $kC_k(\infty)$ for $(a,s,\lambda) =(0,-1,1.25)$ and $N=10^6$.} \label{Fig:diagr}
\end{figure}

We have performed MC simulations, using the approach of \cite{Gillespie}, and observed that for $N \gg 1$ the MC
results for $\langle C_k(t) \rangle$ coincide with predictions of rate equations outside the domain,  associated with
the SCSs, see Fig.~\ref{Fig:C1NCK}b. In the latter  domain, however, the final number of clusters cannot be predicted
by rate equations.   We have observed a sub-linear scaling: $\langle C_k(\infty) \rangle \sim N^{\gamma}$ and $\langle
C(\infty) \rangle \sim N^{\delta}$ with $\gamma, \, \delta <1$, see Fig.~\ref{Fig:diagr}. The non-extensive behavior of
these quantities explains the vanishing densities: $c_k (\infty) \sim N^{-(1-\gamma)}$ and $c(\infty) \sim
N^{-(1-\delta)}$ in the thermodynamic limit. This enigmatic transition from extensive to the observed non-extensive
behavior is caused by fluctuations. To gain analytical understanding, we employ the van Kampen expansion
\cite{vanKampen,Krapivsky}
\begin{equation}
\label{Ck:vK} C_k(t) = N c_k(t) + \sqrt{N} \eta_k(t).
\end{equation}
The terms linear in $N$ are deterministic, and the densities $c_k(t)$ obey \eqref{eq:Gen}. The terms proportional to
$\sqrt{N}$ are stochastic, $\eta_k(t)$ are random variables. To proceed we consider the most simple SCS at
$(a,s,\lambda)=(1,0,1)$ for which a complete analytical solution is available. Using reaction rules \eqref{MC}
we deduce equations for the averages
\begin{subequations}
\label{Ik-eqs}
\begin{align}
\label{I1-eq}
&N\,\frac{d \langle C_1\rangle}{dt} = - 2 \langle C_1(C_1-1)\rangle \\
\label{I2-eq}
&N\,\frac{d \langle C_2\rangle}{dt} =  \langle C_1(C_1-1)\rangle - 3 \langle C_1 C_2\rangle \\
\label{Ik-eq}
&N\,\frac{d \langle C_k\rangle}{dt} =  (k-1)\langle C_1 C_{k-1}\rangle - (k+1) \langle C_1 C_k\rangle
\end{align}
\end{subequations}
with \eqref{Ik-eq} valid for $k\geq 3$. Equations \eqref{Ik-eqs} involve $\langle C_1 C_k\rangle$ with $k\geq 1$. The simplest such quantity, $\langle C_1^2\rangle$, obeys
\begin{equation}
\label{MM-eq}
N\,\frac{d \langle C_1^2\rangle}{dt}= 6 \langle C_1^2\rangle- 4 \langle C_1\rangle - 4 \langle
C_1^3\rangle +W_1
\end{equation}
where $W_{1} = \sum_{k\geq 1} k(k+1) \langle C_1 C_k\rangle$. One finds  $\langle
\eta_k\rangle=0$ for all $k$, see \cite{SM}. Hence $\langle C_1\rangle  = N c_1$ and
\begin{subequations}
\label{M}
\begin{align}
\label{M-av} \langle C_1^2\rangle &=N^2 c_1^2 +N V_1 \\
\label{C3-av} \langle C_1^3\rangle  &= N^3 c_1^3 + 3N^2 c_1 V_1 + N^{3/2}  \langle\eta_1^3 \rangle \\
\label{Ck-av} \langle C_1 C_k\rangle  &= N^2 c_1c_k + N \langle\eta_1 \eta_k  \rangle
\end{align}
\end{subequations}
where $V_1 = \langle\eta_1^2 \rangle=[ \langle C_1^2\rangle - \langle C_1\rangle^2]/N$. Using Eqs.~\eqref{I1-eq} and \eqref{MM-eq} together with expansions \eqref{M} we deduce
\begin{equation}
\label{V1-eq}
\dot{V}_1 + 8V_1 = \sum_{k\geq 1} k(k+1) c_k + 2 c_1 = 2e^\tau+2e^{-2\tau}
\end{equation}
from which $V_1 = \tfrac{2}{9}e^\tau+\tfrac{1}{3}e^{-2\tau}-\tfrac{5}{9}e^{-8\tau}$, or
\begin{equation}
\label{V1-sol}
V_1 =  \tfrac{2}{9}\sqrt{1+2t}+\tfrac{1}{3}(1+2t)^{-1}- \tfrac{5}{9}(1+2t)^{-4}
\end{equation}
in the physical time. Thus fluctuations diverge, and we propose the definition of SCSs, based on this, most
prominent property: SCS is a state where characteristics of a system (clusters number), associated with fluctuations,
prevail over their deterministic counterparts; the characteristics scale sub-linearly with the system size, leading to
vanishing densities (cluster densities) in the thermodynamic limit. The total number of monomers
\begin{equation}
\label{C1:vK}
C_1(\tau) = N e^{-2\tau} + \sqrt{N} \eta_1(\tau)
\end{equation}
exhibits mostly deterministic decay as long as the deterministic part greatly exceeds the stochastic part. Since $V_1=\langle \eta_1^2\rangle \simeq \frac{2}{9}e^\tau$ for $\tau\gg 1$, the stochastic part scales as $\sqrt{N}\,\sqrt{V_1}\sim \sqrt{N}\,e^{\tau/2}$. At time $\tau_*$, when the deterministic part becomes comparable
with the stochastic part,
\begin{equation}
\label{critau} N e^{-2\tau_*}\sim \sqrt{N} e^{\tau_*/2}
\end{equation}
the system enters the SCS. Using \eqref{critau} and $2t=e^{2\tau}-1$ we obtain an estimate of the time when the SCS emerges
\begin{equation}
\label{time:SCS}
 t_* \sim N^{2/5}
\end{equation}
supported by simulations (Fig.~\ref{Fig:CkSCS}a).  At $t>t_*$ the system resides in the SCS where the van Kampen
expansion fails.

Simulations show that after entering the SCS, the system quickly reaches the final stationary state  with vanishing
number of monomers, $C_1=0$, see SM. Thus $\langle C_k (\infty) \rangle  \simeq \langle C_k (t_*)\rangle$ for $k\geq
2$. This allows to estimate the final cluster distribution in the SCS from the cross-over time \eqref{time:SCS} and the
deterministic distribution \eqref{10:crit}, written in the  scaling form as $c_k\simeq (2t)^{-1} e^{-k/\sqrt{2t}}$ and
$c\simeq (2t)^{-1/2}$. Using  $\langle C_k \rangle \sim Nc_k$ and $t \sim t_*$ we find
\begin{equation}
\label{clusters-av:final} 
\langle C_k (\infty) \rangle  \sim N^{3/5}e^{-bk/N^{1/5}}, 
\quad C(\infty) \sim   N^{4/5}
\end{equation}
which agrees with simulations (Fig.~\ref{Fig:CkSCS}b).
\begin{figure}
\centering
\includegraphics[width=4.2cm]{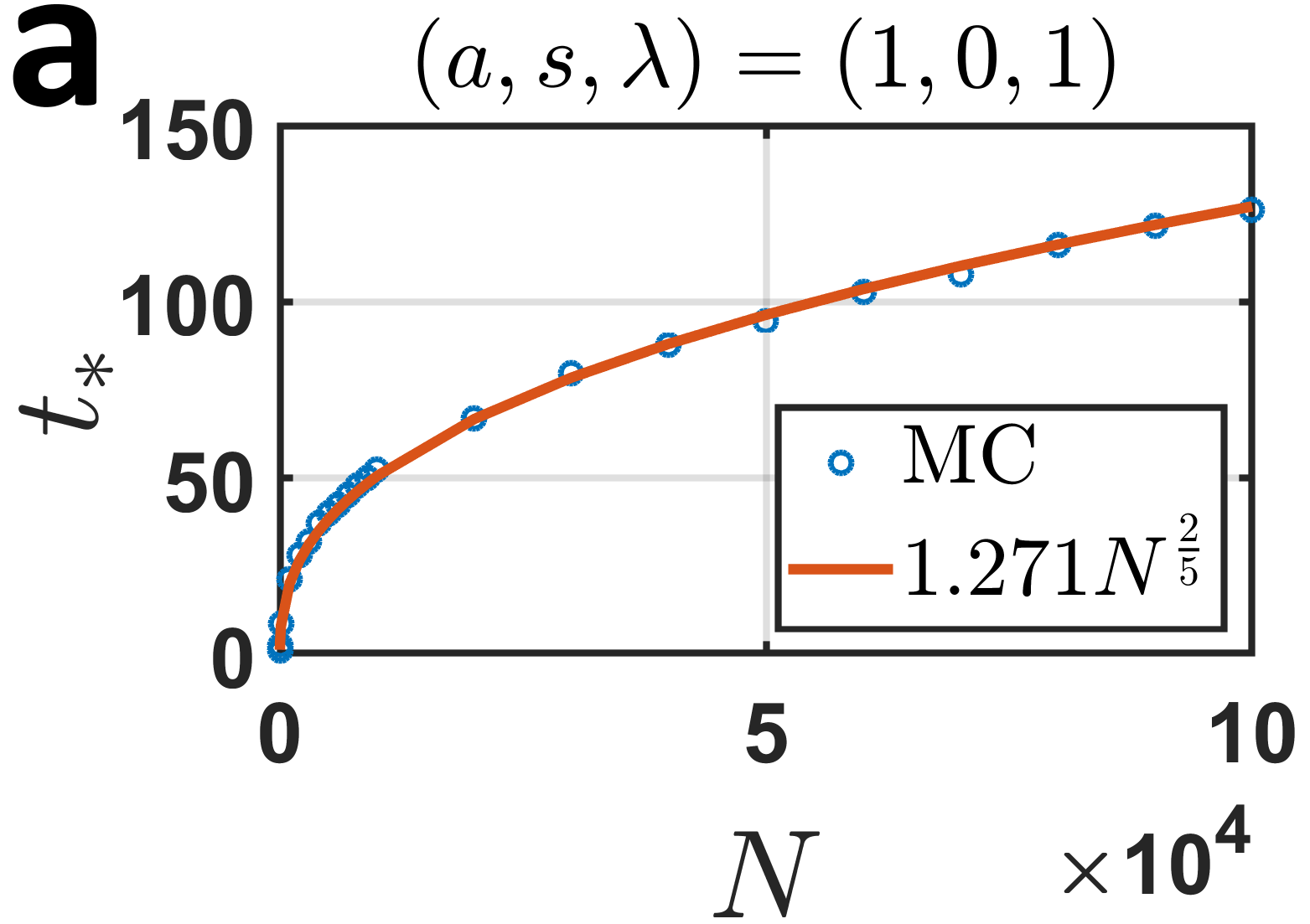}\,
\includegraphics[width=4.2cm]{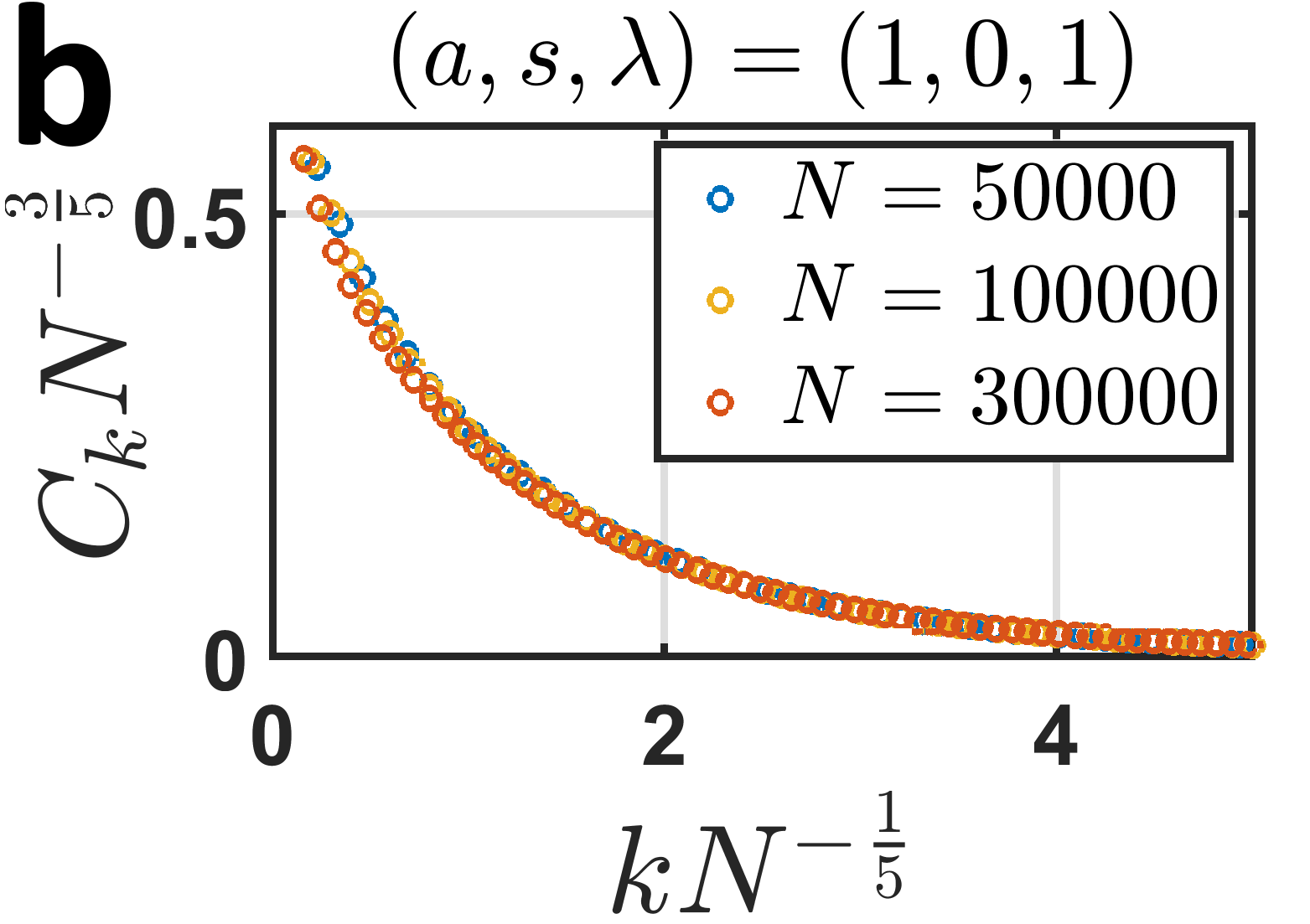}
\caption{(a) Crossover time $t_*$ as a function of system size $N$. Dots -- MC, line -- theory, Eq. \eqref{time:SCS}.
(b) The final cluster size distribution in the SCS with $(a,s,\lambda) =(1,0,1)$ for different $N$. The data collapse
of $C_k(\infty)/N^{3/5}$ on the scaling function $\Phi(\kappa) \sim e^{-b \kappa}$, where $\kappa= k/N^{1/5}$ is
observed, $b \approx 0.87$. }
 \label{Fig:CkSCS}
\end{figure}
The non-extensive growth has been detected in a few aggregation-fragmentation processes with standard spontaneous
fragmentation \cite{bk08,JCP09}, pure aggregation \cite{KrapJam1} and pure fragmentation \cite{KrapJam2}. Neither
monomers nor fluctuations play any special role there. In contrast, the SCSs arising in our models are determined by
fluctuations.

To summarize, the systems undergoing addition and shattering may fall into a  non-extensive state that combines
properties of critical and gelling states. As in a critical state, fluctuations play a dominant role; similar to a
gelling state, mass is mostly accumulated in huge clusters. In the parameter space, the SCS-related domain is
surrounded by standard extensive states, viz. equilibrium and jammed states. The transitions between SCS and ES or JS
are continuous. Our findings demonstrate that a new approach is needed to describe the SCS, which is beyond the van
Kampen expansion. The final cluster distribution is characterized by the exponents $\alpha$, $\beta$, $\gamma$:
\begin{equation}
\label{av-scaling:final} \langle C_k\rangle  \simeq N^{\gamma}\Phi(\kappa), \qquad \kappa= k N^{-\alpha}, \qquad t_*
\sim N^{\beta}.
\end{equation}
The total number of clusters scales as $N^{\delta}$ with $\delta=\gamma+\alpha$. Additionally, $\gamma+2\alpha =1$, due
to mass conservation.

The formation of the SCSs is fluctuation-dominated, so the theoretical understanding is challenging even in the simplest cases. Non-extensive SCSs may influence the operating of large networks where SCSs similar to the reported JSs \cite{Jam3} could possibly emerge.


\newpage
\clearpage

\onecolumngrid

\begin{center}
\Large{\textit{Supplementary material for} \\\textbf{Non-extensive super-cluster states in aggregation with fragmentation}}
\end{center}

\setcounter{equation}{0}
\setcounter{figure}{0}
\setcounter{table}{0}
\setcounter{page}{1}
\renewcommand{\theequation}{S\arabic{equation}}
\renewcommand{\thefigure}{S\arabic{figure}}
\renewcommand{\bibnumfmt}[1]{[S#1]}
\renewcommand{\citenumfont}[1]{S#1}

\maketitle

\vspace{1cm}
\noindent Referring to the equations and figures of the main text, we use bold font.

\section{Solution of rate equations }
\subsection{Model $(a,s)=(1,0)$}

For the model with $(a,s)=(1,0)$,  Eq.~\textbf{[6a]} for the monomer density (recall that we choose the units
where the mass conservation reads $M_1 = 1$) 
\begin{equation}
\label{a1s0c1t} \frac{dc_{1}}{dt} = -(1+\lambda)c_{1}^{2} - c_{1} + \lambda c_{1}
\end{equation}
is a solvable Bernoulli equation which leads to 
\begin{equation}
\label{a1s0c1tsol}
c_1(t)= \frac{1-\lambda}{2 e^{(1-\lambda)t}-1-\lambda}.
\end{equation}
if $c_1(0)=1$. The modified time is defined through
\begin{equation} \label{mod_time}
\tau =\int_0^t c_1(t^{\prime}) dt^{\prime}
\end{equation}
and when the monomer density is given by \eqref{a1s0c1tsol} the modified time is
\begin{equation} \label{a1s0tau}
\tau  = \frac{1}{1+\lambda} \ln \bigg( \frac{2- (1+\lambda)e^{-(1-\lambda )t }}{1-\lambda }\bigg).
\end{equation}
The maximal modified time $\tau_{max}$ corresponds to $t\to \infty$, so it takes the form
\begin{equation}
\label{eq:taumaxlam} \tau_\text{max}(\lambda)=\frac{1}{1+\lambda}\,\ln\frac{2}{1-\lambda}\,.
\end{equation}

Equation \eqref{a1s0c1tsol} shows that in the sub-critical region ($\lambda<1$) the monomer density  decays
exponentially as $t\to\infty$, see  Fig.~\ref{Fig:C1_tau} where subcritical, critical and supercritical evolution of
the monomer density is plotted.

Other cluster densities for $\lambda <1$ saturate at positive values: $\lim_{t\to\infty}c_k(t; \lambda)= c_k(\infty)>0$
for all $k\geq 2$ with $c_k(\infty)$ depending on the initial conditions.  We focus on the mono-disperse initial
conditions: $c_k(0)=\delta_{k,1}$.  Substituting $c_1(\tau)$ from Eq.~\textbf{[7]} into Eq.~\textbf{[6b]} for $k=2$
yields
\begin{equation*}
c_2(\tau)= \frac{\lambda-1}{(1+\lambda)(2+\lambda)} + \frac{2}{1+\lambda}\, e^{-(1+\lambda)\tau}  -
\frac{3}{2+\lambda}\, e^{-(2+\lambda)\tau}.
\end{equation*}
Similarly, using the above result for $c_2(\tau)$ in  Eq.~\textbf{[6b]} with $k=3$, one finds
\begin{eqnarray*}
c_3(\tau) = \frac{2(\lambda-1)}{(1+\lambda)(2+\lambda)(3+\lambda)} + \frac{2}{1+\lambda}\, e^{-(1+\lambda)\tau}
      - \frac{6}{2+\lambda}\, e^{-(2+\lambda)\tau} + \frac{4}{3+\lambda}\, e^{-(3+\lambda)\tau}.
\end{eqnarray*}
Proceeding along the same lines we arrive at
\begin{eqnarray*}
c_k = (\lambda-1)\,\frac{\Gamma(k)\,\Gamma(1+\lambda)}{\Gamma(k+1+\lambda)}+ \frac{2}{1+\lambda}\, e^{-(1+\lambda)\tau}
 - \frac{3(k-1)}{2+\lambda}\, e^{-(2+\lambda)\tau} + \ldots
\end{eqnarray*}
leading at $t \to \infty$ (i.e. at $\tau= \tau_{max}$) to
\begin{eqnarray}
c_k(\infty) =(1-\lambda)\left[\frac{1}{1+\lambda} - \frac{\Gamma(k)\,\Gamma(1+\lambda)}{\Gamma(k+1+\lambda)}\right]
 - \frac{3(k-1)}{2+\lambda}\left[\frac{1-\lambda}{2}\right]^{\frac{2+\lambda}{1+\lambda}} + \ldots
\end{eqnarray}
In the proximity of the critical point  ($0<1-\lambda\ll 1$) the above equation takes the form of Eq.~\textbf{[8]}.

\begin{figure}
\centering
\includegraphics[width=9.5cm]{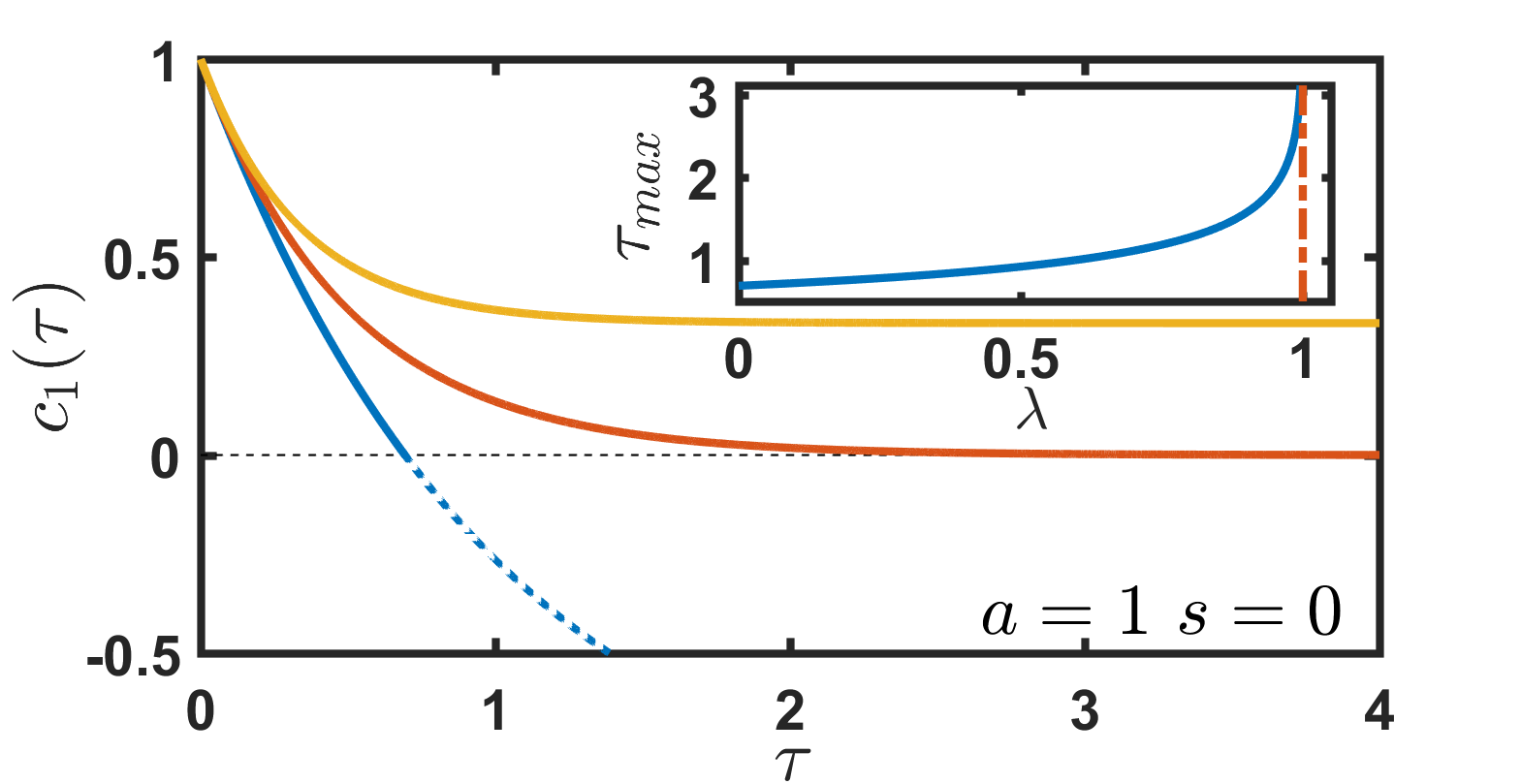}
\caption{The evolution of the monomer density for the model $(a,s)=(1,0)$ and mono-disperse initial conditions. Bottom
to top: sub-critical ($\lambda = 0$), critical ($\lambda =1$), and super-critical ($\lambda=1.19$) behaviors
illustrating evolution to jammed, super-cluster and equilibrium state.  Inset: $\tau_\text{max}$ is an increasing
function of $\lambda$. }
 \label{Fig:C1_tau}
\end{figure}

The total cluster density $N=\sum_{k\geq 1}c_k$ satisfies the rate  equation $\dot N=-\lambda N + 1-\lambda$
from which
\begin{equation}
\label{M0:10} N = 1 - \lambda^{-1}+ \lambda^{-1} e^{-\lambda\tau}.
\end{equation}
When $\lambda =1$, the monomer density satisfies $\dot{c}_1 =-2c_1$, from which $c_1=e^{-2 \tau}$
(for the mono-disperse initial conditions). Using $t= \int_0^{\tau} d\tau^{\prime}/c_1(\tau^{\prime})$ we find
\begin{equation}
\label{tau_t_1_0} \tau =\ln (1+2 t)/2.
\end{equation}
Hence the total cluster density $N=e^{-\tau}$ and the monomer density $c_1=e^{-2 \tau}$ are
\be
\lb{eq:lam1}
c_1(t)=1/(1+2t), \qq \qq N(t) =1/\sqrt{1+2t}
\ee
in terms of the physical time. To find the other densities, we substitute $c_1(\tau)=e^{-2\tau}$ in Eq.~\textbf{[6b]} and find
\begin{equation}
\label{c2c3tau} c_2(\tau)=e^{-2\tau}-e^{-3 \tau}, \qquad c_3(\tau)=e^{-2\tau}-2e^{-3 \tau} +e^{-4\tau} \ldots
\end{equation}
with $\tau$ given by \eqref{tau_t_1_0}. In this way we arrive at Eq.~\textbf{[9]}.

\vspace{0.5cm}

 For the super-critical system $\lambda>1$ the maximal modified time is not limited, $\tau_{max} \to
\infty$ for $t \to \infty$. Hence one can apply the Laplace transform
$$
\widehat{c}_k(p)=\int_0^\infty d\tau \, e^{-p\tau} c_k(\tau),
$$
to Eq.~\textbf{[6b]} to yield

\be \lb{eq:LTck} \widehat{c_k}(p)(p+k+\lambda) = (k-1) \widehat{c_{}}_{k-1}(p). \ee
Solving this equation recursively we get

\be \lb{eq:CkC1} \widehat{c_k}(p)=\frac{(1+p+\lambda)  \Gamma(k) \Gamma(1+\lambda)}{\Gamma(k+1+\lambda +p)}
\widehat{c_1}(p), \ee
where the Laplace transform of the monomer density follows from the Laplace transform of Eq.~\textbf{[6a]}
$$
\widehat{c_1}(p) = \frac{\lambda-1+p}{(\lambda+1+p)p}= \frac{\lambda-1}{\lambda+1}
\frac{1}{p}+\frac{2}{(1+\lambda)(1+\lambda+p)}.
$$
Hence for mono-disperse initial conditions
\be \lb{eq:Ckp} \widehat{c_k}(p)=\frac{(1+p+\lambda)  \Gamma(k) \Gamma(1+\lambda)}{\Gamma(k+1+\lambda +p)}
\left[ \frac{\lambda-1}{\lambda+1} \frac{1}{p}+\frac{2}{(1+\lambda)(1+\lambda+p)} \right]. \ee
For $p\to 0$ the above expression may be written as

\beq \lb{eq:ckpexp}\widehat{c_k}(p) \simeq (\lambda-1)\,\frac{\Gamma(k)\,\Gamma(1+\lambda)}{\Gamma(k+1+\lambda)}
\frac{1}{p}  +\frac{2 \Gamma(k)\,\Gamma(1+\lambda)}{\Gamma(k+1+\lambda)\psi(k+1+\lambda)}\, \frac{1}{ p
+1/\psi(k+1+\lambda) } + \ldots ,  \eeq
where $\psi(x)$ is the digamma function. The first term in Eq.~\eqref{eq:ckpexp} has for $p \to 0$ (that is for $\tau
\to \infty$)  a simple pole $\widehat{c_k}(p) \sim 1/p$, demonstrating the approach to the stationary distribution
 \textbf{[11]}. The relaxation to this stationary distribution is described by the second term in Eq.~\eqref{eq:ckpexp};
it is exponential:

\be \lb{eq:Ck} c_k(\tau) = (\lambda-1)\,\frac{\Gamma(k)\,\Gamma(1+\lambda)}{\Gamma(k+1+\lambda)} + \frac{2
\Gamma(k)\,\Gamma(1+\lambda)}{\Gamma(k+1+\lambda)\psi(k+1+\lambda)} e^{-\tau/\psi(k+1+\lambda)} + \ldots,  \ee
when $\tau \gg 1$.

\vspace{0.5cm}

Note that the transition at $\lambda=1$ from an equilibrium  state (ES) to a stationary jammed state (JS), where final
cluster densities $c_k(\infty)$ depend on initial conditions is essentially a jamming transition. Commonly, the jamming
transition is called the transition, when a variation of some parameter of a system transforms the system from any
other state (e.g. an ES for a spin system or a flowing state in a granular system) into the jammed state which is
stationary and lacks evolution.  For our systems one can state that the jamming transition occurs, when the parameter
$\lambda$  drops below  $2-a$ for a general $a$, see Eq.~\textbf{[18]} of the main text; the physical nature of this
transition is however different. For instance, jamming transition in granular matter occurs when the shear stress drops
down (or packing increases). The system then quickly sets into a stationary JS without any flux. The JS configuration
-- the structure of the system,  will depend on the initial conditions -- its structure at the instant when the stress
drops. In our system the jamming transition manifests by the arrival at a stationary JS with the structure (i.e.
$c_k(\infty)$) that depends on the initial conditions. Hence in spite of the difference in nature, the general features
of JSs and jamming transition are the same in our system and other systems undergoing such transition.

\subsection{Asymptotic analysis}

Applying the Laplace transform to Eq.~\textbf{[12a]}, we obtain for mono-disperse initial conditions,
$\widehat{c}_{k}(p)=\widehat{c}_{k-1}(p)/(p+1+\lambda k^{-1})$, which is iterated to find
\begin{equation}
\widehat{c}_{k}(p) = \frac{(1+\lambda \epsilon)\Gamma(1+\lambda \epsilon) k!}{\Gamma(k+1+\lambda \epsilon)}
\epsilon^{k-1} \widehat{c}_{1}(p) \label{eq:ckpc1p}
\end{equation}
where $\epsilon  = (1+p)^{-1}$. Applying the Laplace transform to the mass density $\sum_{k \geq 1} k c_{k}(\tau) =1$
gives $\sum_{k \geq 1} k \widehat{c}_{k} =1/p$. Plugging into this sum $\widehat{c}_{k}(p)$ given by \eqref{eq:ckpc1p}
and expressing the sum through the hypergeometric function,
$$
\sum_{k=1}^{\infty} \frac{k! \epsilon^k}{\Gamma(k+1+\lambda \epsilon)} = \frac{\epsilon F\left[2,2;2+\lambda \epsilon;
\epsilon\right]}{\Gamma(2+\lambda \epsilon)},
$$
yields
\begin{equation}
\label{eq:c1pgen}
 p\widehat{c}_{1}(p) =\frac{1}{F\left[2,2;2+\lambda\epsilon; \epsilon\right]}.
\end{equation}
Combining \eqref{eq:ckpc1p} and \eqref{eq:c1pgen} we arrive at Eq.~\textbf{[14]}.

The long time behaviors can be extracted from the $p \to 0$ behavior of the corresponding Laplace transforms.
Specializing the integral representation of the hypergeometric function
$$
F\left[a,b; c;z \right] = \frac{\Gamma(c)}{\Gamma(b) \Gamma(c-b)} \int_0^1 dx \frac{x^b (1-x)^{c-b-1}}{(1-xz)^a}
$$
to $a=2$, $b=2$, $c=2+\lambda\epsilon$ and $z=\epsilon$ we establish Eq.~\textbf{[15]}. Since $\epsilon=(1+p)^{-1}$,
for $p \to 0$ we replace $\epsilon$ by $1$ in the right-hand side of Eq.~\textbf{[15]}, apart from the denominator
$(1-x \epsilon)^{-2}$ where one should be more careful. Writing $1-x \epsilon =1-x + x(1- \epsilon)$ and analyzing the
integral we find that its dominant part is gathered in the region $1-x = {\cal O} (1-\epsilon)$. Since $(1-\epsilon)
\simeq p$ as $p \to 0$, we write $1-x =py$ to recast Eq.~\textbf{[15]} to
$$
F\left[2,2; 2 +\lambda \epsilon; \epsilon \right] \simeq \lambda (1+\lambda) p^{\lambda-2} \int_0^{\infty} dy
\frac{y^{\lambda-1}}{(1+y)^2}.
$$
Computing the integral we finally arrive at
\begin{equation}
\label{eq:F222}
 F\left[2,2; 2 +\lambda \epsilon; \epsilon \right] \simeq \lambda (1+\lambda) \frac{\pi (1-\lambda)}{
\sin( \pi \lambda)}p^{\lambda-2} .
\end{equation}
By inserting \eqref{eq:F222} into \eqref{eq:c1pgen} we find
\begin{equation}
\label{eq:c1ppto0} \widehat{c}_{1}(p) \simeq \frac{\sin( \pi \lambda)}{\pi (1-\lambda)}
\frac{p^{1-\lambda}}{\lambda(1+\lambda)}
\end{equation}
 for $p \to 0$, from which we deduce the $\tau \to \infty$ asymptotic
\begin{equation}
\label{c1:0-1-decay} c_1(\tau) \simeq -  \frac{\sin( \pi \lambda)}{\pi \Gamma(2+\lambda)} \frac{1}{\tau^{2-\lambda} }.
\end{equation}
To extract the asymptotic decay of the total cluster density is to use Eq.~\textbf{[12a]} and \eqref{c1:0-1-decay} to
conclude that $c\simeq \frac{\lambda+1}{\lambda-1}\,c_1$. Therefore
\begin{equation}
\label{N:0-1-decay} c(\tau)\simeq \frac{\sin[\pi(\lambda-1)]}{\pi
(\lambda-1)}\,\frac{1}{\Gamma(\lambda+1)}\,\frac{1}{\tau^{2-\lambda}}.
\end{equation}
Using $t=\int_0^\tau d\tau'/c_1(\tau')$, that follows from the definition of $\tau(t)$, we find $t= -\sin \pi \lambda
/[\pi \Gamma(2+\lambda)] \tau^{3-\lambda}$ which yields
\begin{subequations}
\begin{align}
&c_1 = A(\lambda)\, t^{-\frac{2-\lambda}{3-\lambda}}, \qquad c =  \frac{\lambda+1}{\lambda-1}\,A(\lambda)\,
t^{-\frac{2-\lambda}{3-\lambda}}
\label{c1N:critical}\\
&A(\lambda) = \left[-\frac{\sin(\pi\lambda)}{\pi \Gamma(\lambda+2)}\right]^{\frac{1}{3-\lambda}}
(3-\lambda)^{-\frac{2-\lambda}{3-\lambda}}. \label{A:critical}
\end{align}
\end{subequations}

Consider now the case of $s=a-1$.  From Eqs.~\textbf{[17b]}  we iteratively obtain
\begin{equation}
\label{Ckp_C1p} \widehat{c}_k(p)=\widehat{c}_1(p) \prod_{j=2}^k \frac{(j-1)^a}{j^a +\lambda j^{a-1} +p}.
\end{equation}
Using again the mass conservation, $\sum_{k \geq 1}k\widehat{c}_k(p)=1$, we find that $p\widehat{c}_1(p) =
1/G(p,\lambda)$ with
\begin{equation}
\label{Cp-lambda} G(p,\lambda)=\sum_{k\geq 1} k\prod_{j=2}^k \frac{(j-1)^a}{j^a+\lambda j^{a-1}+p}.
\end{equation}
The final state of $\tau \to \infty$ corresponds to $p \to 0$. Setting $p=0$ on the right-hand side of
\eqref{Cp-lambda} and massaging the sum we obtain
\begin{equation}
\label{C0-lambda1} G(0,\lambda) = \sum_{k\geq 1} k^{1-a}\,\frac{\Gamma(k+1)\Gamma(2+\lambda)}{\Gamma(k+1+\lambda)}.
\end{equation}
The summands behave as $\Gamma(2+\lambda)\,k^{1-a-\lambda}$ when $k\gg 1$, so the sum on the right-hand side of
\eqref{C0-lambda1} converges when $\lambda>2-a$ and $c_1(\infty) =\lim_{p \to 0} p\widehat{c}_1(p)=1/G(0,\lambda)$. For
$\lambda \leq 2-a$ the sum  in \eqref{C0-lambda1} diverge, yielding vanishing final densities. Hence $\lambda_{\rm
up}=2-a$.

The final densities $c_k (\infty)= \lim_{p\to 0} \widehat{c}_k(p)/p$ are found by combining Eqs.~\eqref{Ckp_C1p} and
\eqref{C0-lambda1} with $p\widehat{c}_1(p) = 1/C(p,\lambda)$. This yields Eq.~\textbf{[18]}.

\subsection{Model $(a,s)=(a,a-1)$}

To analyze the relaxation to the final density distribution   \textbf{ [18]} we consider the small $p$ behavior of the
amplitude $G(p,\lambda)$ given by Eq.~\eqref{Cp-lambda}. For $p\to 0$ we write,
\begin{eqnarray}
\label{Cplam} G(p,\lambda) &=& \sum_{k\geq 1} k\prod_{j=2}^k \frac{(j-1)^a}{j^a+\lambda j^{a-1}+p} \simeq \sum_{k\geq
1} k^{1-a}\,\frac{\Gamma(k+1)\Gamma(2+\lambda)}{\Gamma(k+1+\lambda)}
\prod_{j=2}^k\frac{1}{1+\frac{p}{j^a}} \\
&\sim & \int_{1}^\infty \frac{dk}{k^{a+\lambda-1}}\,\Gamma(2+\lambda)\,e^{-pk^{1-a}/(1-a)} .\nonumber
\end{eqnarray}
For $\lambda >2-a$ the limiting value of $G(0, \lambda)$ takes the form of Eq.~\eqref{C0-lambda1}. For $\lambda \leq
2-a$ the sum in Eq.~\eqref{Cplam} diverges at $p=0$, while for small positive $p$ the dominant part is gathered when
$k\sim p^{-1/(1-a)}\gg 1$, and for such large $k$ the replacement of summation by integration is justified. We
emphasize again that the condition $a<1$ is assumed. The result of the above integration depends on the value of
$\lambda$ and yields,

\begin{equation}
\label{eq:Cplam1} G(p,\lambda) \simeq
\begin{cases}
\frac{\Gamma(4-a)}{1-a}\,\ln \frac{1-a}{p} & \lambda  =2-a  \\
\Gamma(2+\lambda)\,\Gamma(1-\gamma)\,(1-a)^{-\gamma}\,p^{\gamma-1} & 1< \lambda <2-a ,
\end{cases}
\end{equation}
where $\gamma \equiv \frac{\lambda-1}{1-a}$. This parameter varies in the range $0<\gamma<1$ in the critical region
$1<\lambda < 2-a$. Using $\widehat{c}_1(p)=1/[pG(p,\lambda)]$ and making the inverse Laplace transform  we extract the
large time asymptotic,
\begin{equation}
\label{eq:Cplam2} c_1 (\tau) \sim
\begin{cases}
\frac{1}{\ln[(1-a) \tau]} & \lambda =2-a  \\
\tau^{-(1-\gamma)} & 1 < \lambda <2 ,
\end{cases}
\end{equation}
or, in physical time,
\begin{equation}
\label{eq:Cplam3} c_1 (t) \sim
\begin{cases}
\frac{1}{\ln[(1-a)^2 t ]} & \lambda =2-a  \\
t^{-\frac{1-\gamma}{2-\gamma} } & 1 < \lambda <2 .
\end{cases}
\end{equation}
Similar analysis may be done to obtain the total number of clusters $c(t)$ and cluster densities $c_k(t)$.

\subsection{Rate equations for finite-size  systems}
 Let us consider the model with the rates
\begin{equation*}
A_k = k, \qquad  \qquad S_k = \lambda
\end{equation*}
Then the standard rate equations, corresponding Eq.~\textbf{[6]} of the main text read
\begin{subequations}
\begin{align}
\dot{c}_1 &= -c_{1} -\sum_{j=1}^{\infty} j c_{j} + \lambda \sum_{j=2}^{\infty} j c_{j} \\
\dot{c}_k&= (k-1)c_{k-1} - k c_k  - \lambda c_{k}, \quad k\geq 2
\end{align}
\end{subequations}
where dot is derivative with respect to $\tau$, see the main text.

Suppose only clusters up to size $N$ can form. What happens if the heaviest cluster of size $N$ is hit by a monomer?
Addition is forbidden as we forbid formation of clusters of size exceeding $N$. Thinking about $\lambda$ as the
relative weight of shattering compared to addition, we should forbid {\em both} addition and shattering of clusters of
size $N$. Then the rate equations read
\begin{subequations}
\label{long}
\begin{align}
\dot{c}_1 &= -c_{1} -\sum_{j=1}^{N-1} j c_{j} + \lambda \sum_{j=2}^{N-1} j c_{j} \\
\dot{c}_k&= (k-1)c_{k-1} - k c_k  - \lambda c_{k}, \quad 2\leq k <N \\
\dot{c}_N&=(N-1)c_{N-1}
\end{align}
\end{subequations}
When $\lambda=1$, these equations simplify to
\begin{subequations}
\label{1}
\begin{align}
\dot{c}_k&= (k-1)c_{k-1} - (k+1) c_k, \quad 1\leq k <N \\
\dot{c}_N&=(N-1)c_{N-1}
\end{align}
\end{subequations}
These equations are recurrent and they are solved to yield
\begin{subequations}
\label{1-sol}
\begin{align}
c_k&= e^{-2\tau}[1-e^{-\tau}]^{k-1}, \quad 1\leq k <N \\
c_N&=N^{-1}[1-e^{-\tau}]^N   +e^{-\tau}[1-e^{-\tau}]^{N-1}
\end{align}
\end{subequations}
When $\tau=\infty$, this solution predicts
\begin{subequations}
\label{1-sol-inf}
\begin{align}
c_k&= 0, \quad 1\leq k <N \\
c_N&=N^{-1}
\end{align}
\end{subequations}
i.e., all mass is engulfed by the single cluster of mass $N$. The conceptual difficulty here is the very peculiar
properties of the largest cluster of size $N$ -- it is completely inert (no addition and no shattering). This seems to
be physically implausible.

Let us consider another rate equations model for finite-size system. Let the addition process $[1] + [N] \to [N+1]$ be
forbidden, while shattering still allowed, then instead of \eqref{long} one gets
\begin{subequations}
\begin{align}
\dot{c}_1 &= -c_{1} -\sum_{j=1}^{N-1} j c_{j} + \lambda \sum_{j=2}^{N} j c_{j} \\
\dot{c}_k&= (k-1)c_{k-1} - k c_k  - \lambda c_{k}, \quad 2\leq k <N \\
\dot{c}_N&=(N-1)c_{N-1} - \lambda c_N
\end{align}
\end{subequations}
which for $\lambda=1$ simplify to
\begin{subequations}
\label{2}
\begin{align}
\dot c_1 & = -2c_1+ Nc_N  \\
\dot{c}_k&= (k-1)c_{k-1} - (k+1) c_k, \quad 2\leq k <N \\
\dot{c}_N&=(N-1)c_{N-1}-c_N
\end{align}
\end{subequations}
Equations \eqref{2} are more conceptually problematic than \eqref{1}. Indeed, if there is a cluster of mass $N$, it
should be a single cluster that engulfed the entire mass (recall that the total mass is $N$). Certainly there should be
no monomers to trigger shattering, that is, the last term in Eq.~\eqref{2}c is spurious. Of course, Eqs.~\eqref{1}  or
\eqref{2} become bad much earlier, for $k\ll N$ (when a single monomer is left), but still Eqs.~\eqref{2} are
conceptually inconsistent.

Let us now look at the stationary solution, which reads,
\begin{subequations}
\label{2solA}
\begin{align}
c_k&= \frac{1}{k(k+1)\,H_N}\,, \quad 1\leq k <N \\
c_N&= \frac{1}{N H_N}
\end{align}
\end{subequations}
where $H_N = \sum_{1\leq k\leq N} k^{-1}$ are harmonic numbers. Again we notice the physical inconsistency: If
$c_N(\infty)$ is not zero, all other concentrations $c_k(\infty)$ with $1 \leq k \leq N-1$ must be zero, as all mass
belongs to the largest cluster. However, this is not the case for Eqs.~\eqref{2solA}.

\section{Super cluster states }

Here we give the detail for the derivation of Eq.~\textbf{[25]}. Writing Eq.~\textbf{[21]} for  monomers we express
$\langle C_1\rangle$ and $\langle C_1^2\rangle$ through $\langle\eta_1\rangle$ and $\langle\eta_1^2\rangle$:
\begin{subequations}
\begin{align}
\label{M-av}
\langle C_1\rangle  &= N c_1 + \sqrt{N} \langle\eta_1\rangle\\
\label{MM-av} \langle C_1^2\rangle  &= N^2 c_1^2 + 2N^{3/2} c_1\langle\eta_1\rangle + N\langle\eta_1^2\rangle
\end{align}
\end{subequations}

Plugging expansions \eqref{M-av}--\eqref{MM-av} into Eq.~\textbf{[22a]} for monomers and equating the leading terms of
the order $O(N^2)$ we recover the rate equation for the density of monomers. Equating the sub-leading terms of the
order $O(N^{3/2})$ yields
\begin{equation}
\label{eta-av:1} \frac{d \langle\eta_1\rangle}{d \tau}= - 4 \langle\eta_1\rangle
\end{equation}
The evolution begins with a deterministic initial state, so $\eta_1(0)=0$, so the solution is trivial:
$\langle\eta_1\rangle = 0$.

Similarly $\langle C_2\rangle  = N c_2 + \sqrt{N} \langle\eta_2\rangle$ and
\begin{equation}
\label{C12-av} \langle C_1 C_2\rangle  = N^2 c_1 c_2 + N^{3/2} c_1\langle\eta_2\rangle + N\langle\eta_1 \eta_2\rangle
\end{equation}
Plugging \eqref{M-av}--\eqref{MM-av} and \eqref{C12-av} into Eq.~\textbf{[22b]}, written for dimers,
\begin{equation}
\label{I2-eq} N\,\frac{d \langle C_2\rangle}{dt} =  \langle C_1(C_1-1)\rangle - 3 \langle C_1 C_2\rangle,
\end{equation}
and equating the leading terms of the order $O(N^2)$ we recover the rate equation $\dot c_2=c_1^2-3c_1c_2$; equating
the sub-leading terms of the order $O(N^{3/2})$ we get
\begin{equation}
\label{eta-av:2} \frac{d \langle\eta_2\rangle}{d \tau}= -3\langle\eta_2\rangle
\end{equation}
Since $\eta_2(0)=0$, the solution is also trivial: $\langle\eta_2\rangle = 0$.

Similarly we use Eq.~\textbf{[22c]} for $k \geq 3$ and recursively establish
\begin{equation}
\label{eta-sol:k} \langle\eta_k\rangle = 0.
\end{equation}
Thus \eqref{eta-sol:k} holds for all $k\geq 1$.

Subtracting Eq.~\textbf{[22a]} for $k =1$, multiplied by $2\langle C_1\rangle$ from Eq.~\textbf{[23]}, we obtain
\begin{eqnarray}
\label{MMV-eq} N\,\frac{d}{dt}\!\left[ \langle C_1^2\rangle- \langle C_1\rangle^2 \right]  = 4 \langle C_1^2\rangle
\langle C_1\rangle - 4 \langle C_1^3\rangle + 2 \langle C_1^2\rangle + \sum_{k\geq 1} k(k+1) \langle C_1 C_k\rangle +
4\left[\langle C_1^2\rangle -  \langle C_1\rangle^2 -  \langle C_1\rangle\right].
\end{eqnarray}

Since $\langle\eta_1\rangle = 0$, equation \eqref{MM-av} becomes
\begin{equation}
\label{V1:def} \langle C_1^2\rangle = N^2c_1^2 + N V_1, \qquad V_1=\langle\eta_1^2\rangle
\end{equation}
More generally
\begin{equation}
\label{V1k:def} \langle C_1 C_k\rangle   = N^2 c_1 c_k + N\langle\eta_1 \eta_k\rangle
\end{equation}
for all $k\geq 1$. We also compute
\begin{equation}
\label{C1C1} \langle C_1^3\rangle=N^3c_1^3+3N^2 c_1V_1+N^{3/2}\langle\eta_1^3\rangle
\end{equation}
Substituting \eqref{V1:def}--\eqref{C1C1} into Eq.~\eqref{MMV-eq} and equating terms in the leading $O(N^2)$ order we
arrive at
\begin{equation}
\label{V1-eq} \frac{d V_1}{d\tau} + 8V_1 = \sum_{k\geq 1} k(k+1) c_k + 2 c_1,
\end{equation}
which is the equation \textbf{[25]} of the main text.

When a system enters SCS at $t\sim t_*$, it undergoes a rather short evolution to the final jammed state. The
 monomer density sharply drops to zero, see Fig.~\ref{Fig:Supll_6}. Hence in the SCS $c_k(\infty) \simeq c_k(t_*)$.
\begin{figure}
\centering
\includegraphics[width=5cm]{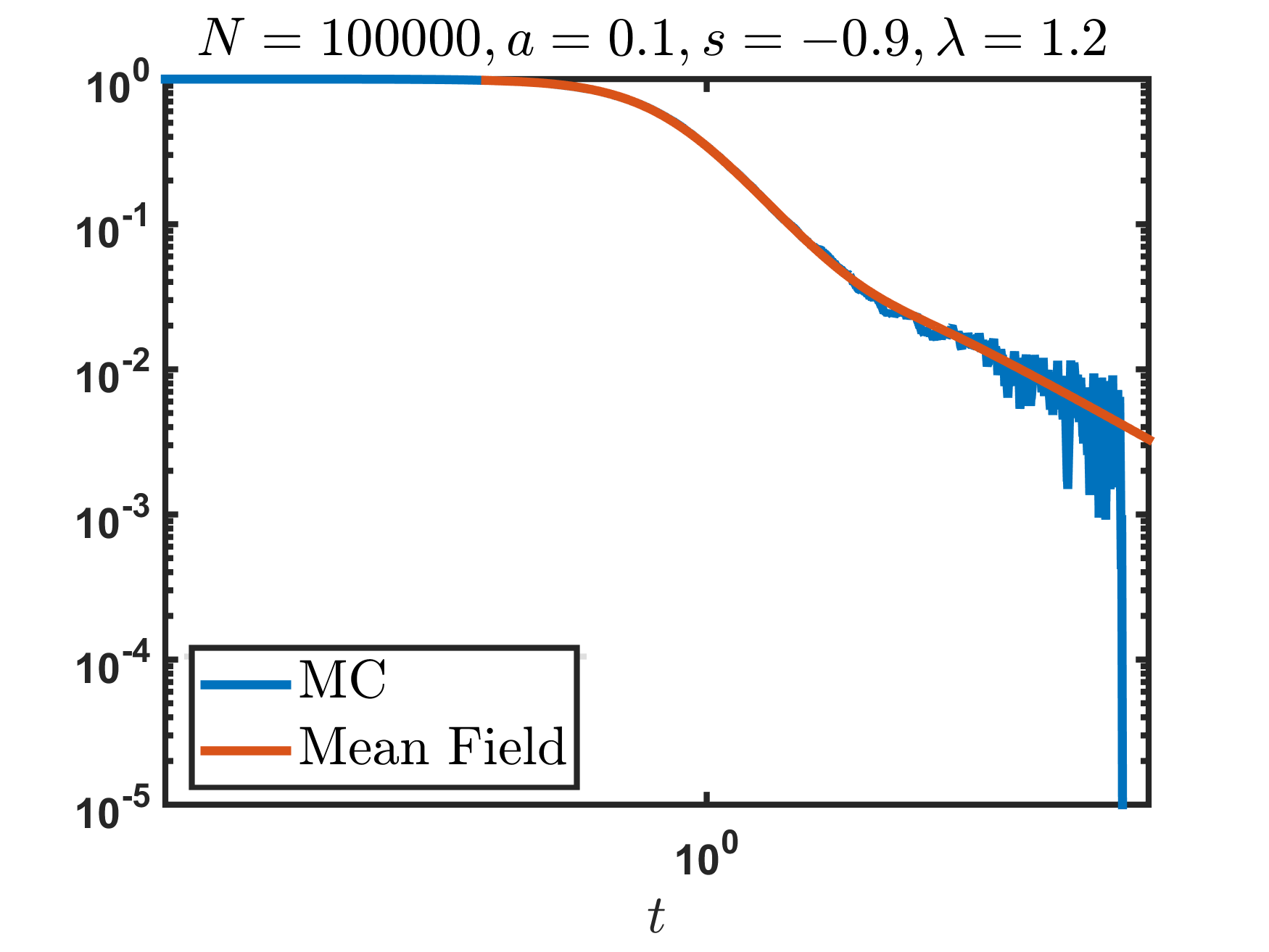}\q
\includegraphics[width=5cm]{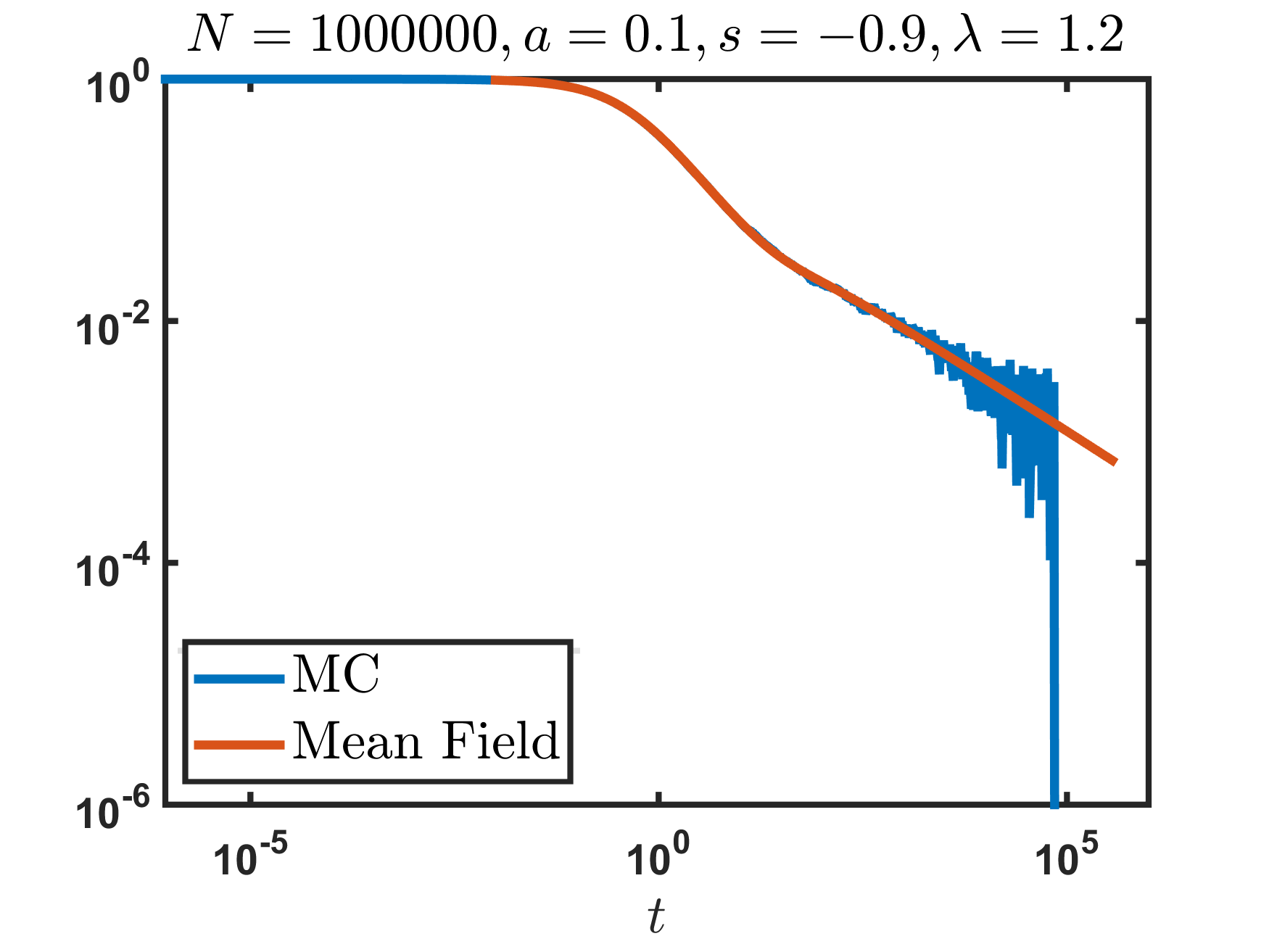}\q
\includegraphics[width=5cm]{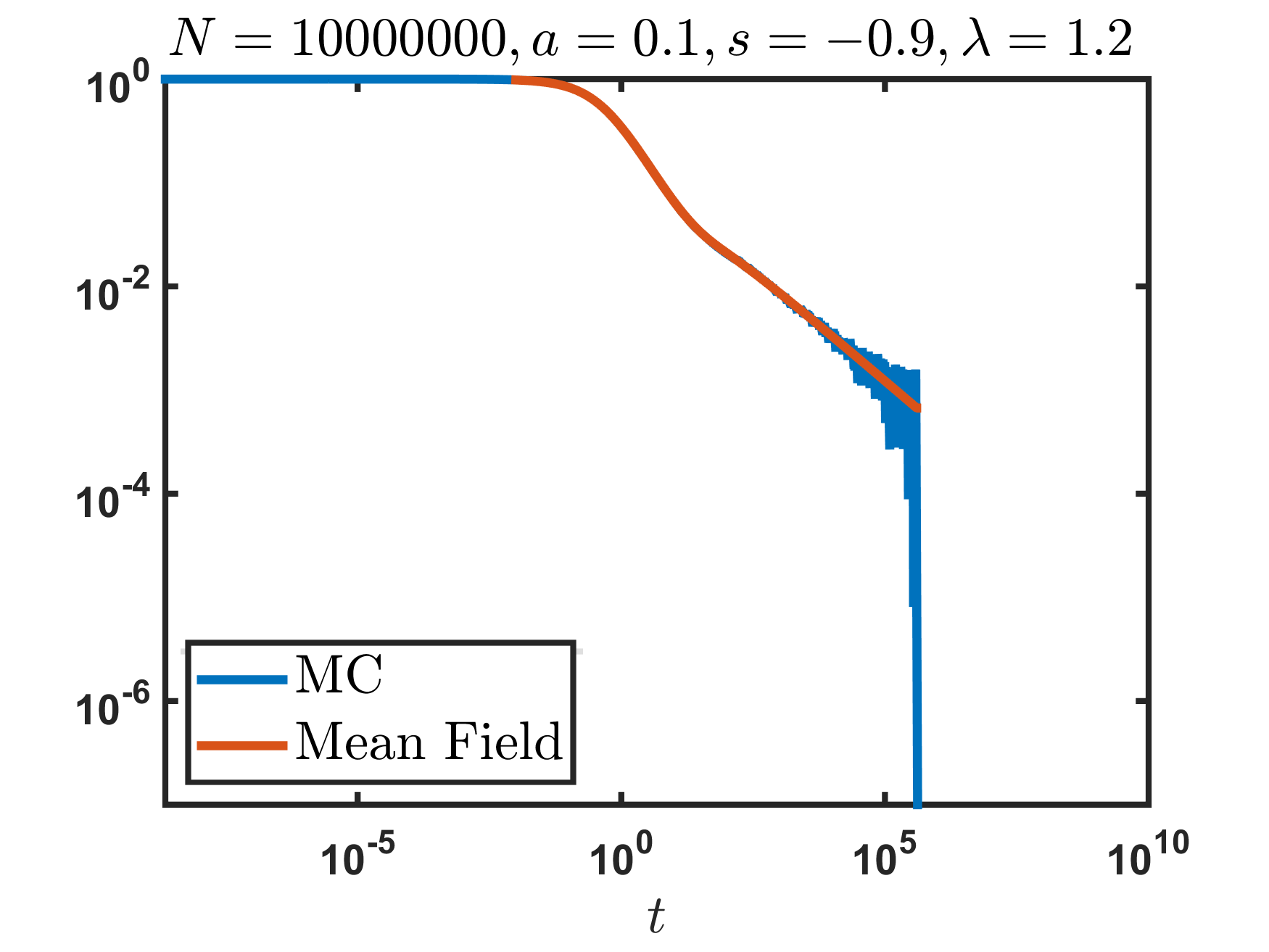}
\caption{Evolution of the monomer density for the parameters $a=0.1$, $s=-0.9$ and $\lambda=1.2$ corresponding to the
SCS for different system size: $N= 10^5$ (left),  $N= 10^6$ (middle), and $N= 10^7$ (right). Shortly after the system
enters the SCS, where the fluctuations dominate, the monomer density sharply drops to zero. }
 \label{Fig:Supll_6}
\end{figure}

\section{Weak phase transitions}

On the boundary of the SCS, $\lambda_{\rm low}=1$, the evolution of the cluster densities $c_k(t)$ undergoes an
infinite series of discontinuous phase transitions. These occur in the thermodynamic limit at the critical values of
the exponent $a$ characterizing the addition rate. The critical values $a_p$ with $p=1,2,3,4,\ldots$ are determined by
$a_1=1$ and then recursively by
\begin{equation}
\label{ap} p^{a_p}+p^{a_p-1} = (p-1)^{a_p}+(p-1)^{a_p-1}.
\end{equation}
These critical values decrease as $p$ increases: $a_2\approx 0.41503749$, $a_3\approx 0.29048870$, $a_4\approx
0.22433973$, etc. and approach to zero according to
\begin{equation*}
a_p=\frac{1}{p} - \frac{1}{2p^2}+\frac{5}{12p^3} - \frac{7}{24 p^4} + \ldots
\end{equation*}
when $p\to\infty$.

To demonstrate this, we start with the evolution of monomers, $c_1=e^{-2\tau}$ and solve the rate equation for dimers
$\frac{dc_2}{d\tau}+(2^a+2^{a-1})c_2=c_1$, which yields,
\begin{equation}
\label{c2:a} c_2 = \frac{e^{-2\tau}-e^{-(2^a+2^{a-1})\tau}}{2^a+2^{a-1}-2}.
\end{equation}
This equation shows that if $2^a+2^{a-1}>2$, or equivalently $a>a_2$ with $a_p$ defined by \eqref{ap} for $p=2$, the
dimer density decays similarly to the monomer density, that is,  $ c_2 \sim e^{-2\tau}$. All densities actually decay
similarly,
\begin{equation}
c_k\simeq {\cal B}_k e^{-2\tau}\quad\text{for}\quad k\geq 2,
\end{equation}
or, in physical time for  $t \gg 1$,
\begin{equation} \label{cka1}
c_k \simeq \frac{{\cal B}_k}{t} \quad\text{for}\quad k\geq 2,
\end{equation}
where we use $2\tau = \ln(1+2t)$. The amplitudes may be found recursively:
\begin{equation}
\mathcal{B}_k = \prod_{j=2}^k \frac{(j-1)^a}{j^a + j^{a-1}-2}.
\end{equation}
We estimate the above amplitudes $\mathcal{B}_k$ as
\begin{eqnarray}
\mathcal{B}_k \eq \prod_{j=1}^k \frac{(1-1/j)^a}{ (1+1/j -2/j^a)} = \prod_{j=2}^k \left(1-\frac{1}{j} \right)^a \cdot
\exp\left[ -
\sum_{j=2}^k \log \left(1+\frac{1}{j}-\frac{2}{j^a} \right) \right] \\
&\sim & \frac{1}{k^a} \exp\left[ - \int_2^k \log \left(1+\frac1j - \frac{2}{j^a} \right) dj \right] \approx
\frac{1}{k^a} \exp\left[- \int_2^k \frac{dj}{j}
+2 \int_2^k \frac{dj}{j^a}\right] \nonumber  \\
& \sim & \frac{2}{k^{1+a} } e^{2k^{1-a}/(1-a)}. \nonumber
\end{eqnarray}
Conservation of mass then reads,
\begin{eqnarray}
\sum_{k=1}^{ k_{\rm max} } k \mathcal{B}_k e^{-2 \tau}  \approx  e^{-2 \tau} \sum_{k=1}^{ k_{\rm max} } \frac{2}{k^{a}}
e^{2k^{1-a}/(1-a)}  \approx  e^{-2 \tau} \int_1^{k_{\rm max}} \frac{2}{x^a} e^{2 x^{1-a}/(1-a)} dx \simeq  e^{2 k_{\rm
max}^{1-a}/(1-a)-2 \tau}=1 ,\nonumber
\end{eqnarray}
which yields $k_{\rm max} = [(1-a) \tau]^{1/(1-a)}$. Hence we obtain for the total cluster density,
\begin{eqnarray}
c(\tau) &=& \sum_{k=1}^{k_{\rm max}} \mathcal{B}_k e^{-2 \tau}  \simeq  e^{-2 \tau} \sum_{k=1}^{k_{\rm max}}
\frac{2}{k^{1+a}} e^{2k^{1-a}/(1-a)} \\
& \sim  & e^{-2 \tau} \frac{1}{k_{\rm max} } \int_1^{k_{\rm max} } \frac{2}{x^a} e^{2 x^{1-a}/(1-a)} dx =
\frac{1}{k_{\rm max}} = [(1-a) \tau]^{-1/(1-a)} \nonumber .
\end{eqnarray}
Thus when $a_2<a<a_1=1$, the total density behaves  asymptotically in physical time as
\begin{equation}
\label{Nt:2} c(t) \sim  \frac{1}{\left( \ln t  \right)^{1/(1-a)}}.
\end{equation}

When $a=a_2$, the density of dimers becomes
\begin{equation}
\label{c2:a2} c_2 = \tau e^{-2\tau}
\end{equation}
and generally
\begin{equation}
c_k\simeq \mathcal{B}_k \tau e^{-2\tau}\quad\text{for}\quad k\geq 2
\end{equation}
with amplitudes
\begin{equation}
\mathcal{B}_k = \prod_{j=3}^k \frac{(j-1)^{a_2}}{j^{a_2} + j^{a_2-1}-2}.
\end{equation}
Applying the same analysis as before we obtain for the asymptotic behavior of the total cluster density,
\begin{equation}
\label{Nt:2*} c(t)  \sim \frac{1}{ \left[ \ln \left( t / \ln t \right) \right]^{1/(1-a_2)}} .
\end{equation}
Similarly, when $a_3 < a < a_2$, we have
\begin{equation}
\label{cka23} c_k \simeq \mathcal{B}_k e^{-\left(2^a +2^{a-1}\right) \tau}  \quad\text{for}\quad k\geq 2 .
\end{equation}
with amplitudes
\begin{equation}
\label{Bka23} \mathcal{B}_k = \frac{1}{2-2^a - 2^{a-1}} \prod_{j=3}^k \frac{(j-1)^{a}}{j^{a} + j^{a-1}-2^a - 2^{a-1}}.
\end{equation}
In physical time, for  $t \gg 1$,
\begin{equation} \label{cka1}
c_k \simeq \frac{ {\cal B}_k }{ t^{ (2^a + 2^{a-1} )/2}} \quad\text{for}\quad k\geq 2.
\end{equation}
Using the amplitudes \eqref{Bka23} and conservation of mass we obtain Eq.~\eqref{Nt:2} for the total cluster density
$c(t)$ for $a_3 < a < a_2$,  and Eq. \eqref{Nt:2*} for $a=a_3$, with $a_2$ substituted by $a_3$.

Continuing these calculations,   other laws for the asymptotic behavior of $c_k(t)$ and $N(t)$ for $\lambda=1$ may be
obtained. Namely, we find,

\begin{figure}
\centering
\includegraphics[width=7.5cm]{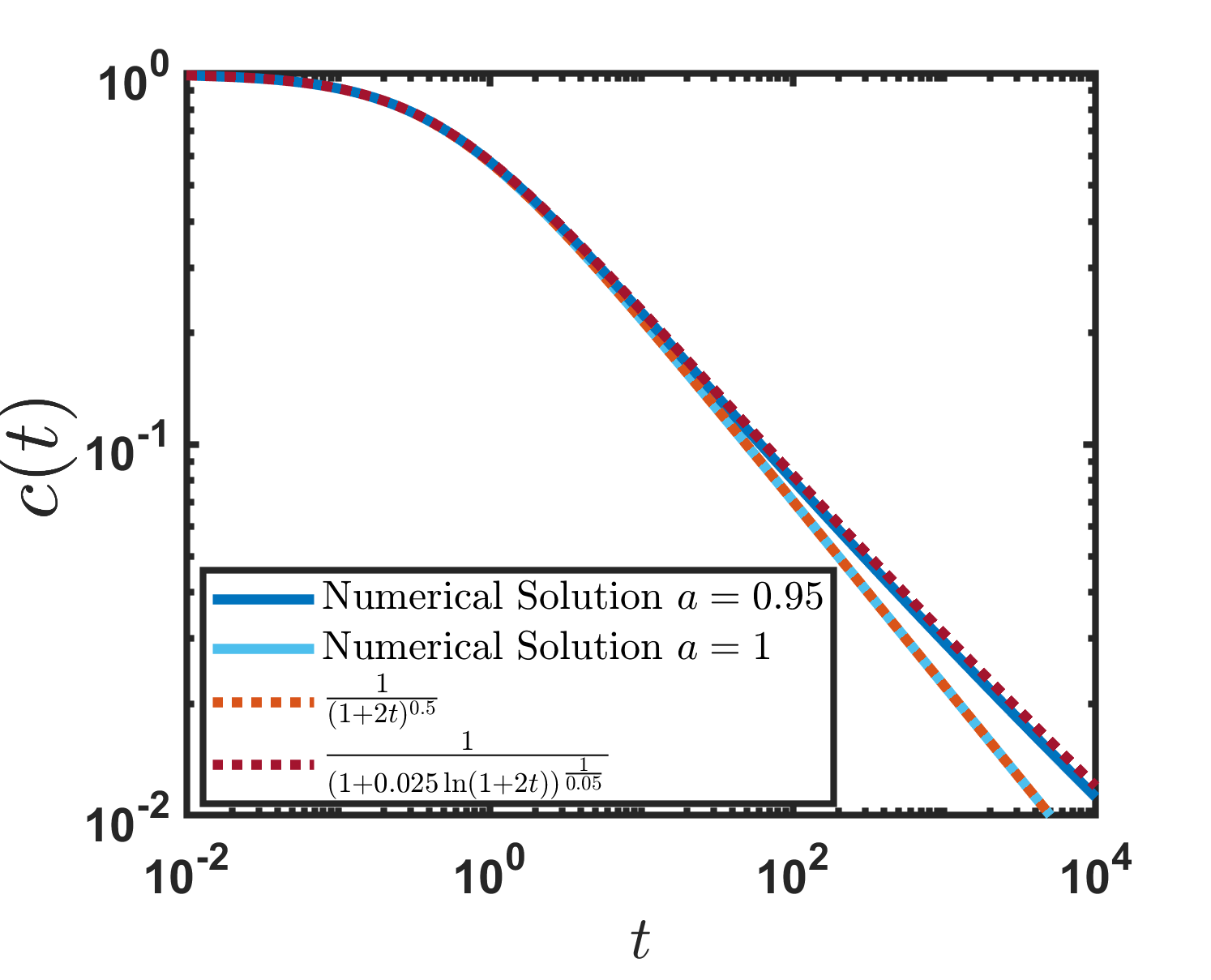}\q
\includegraphics[width=7.5cm]{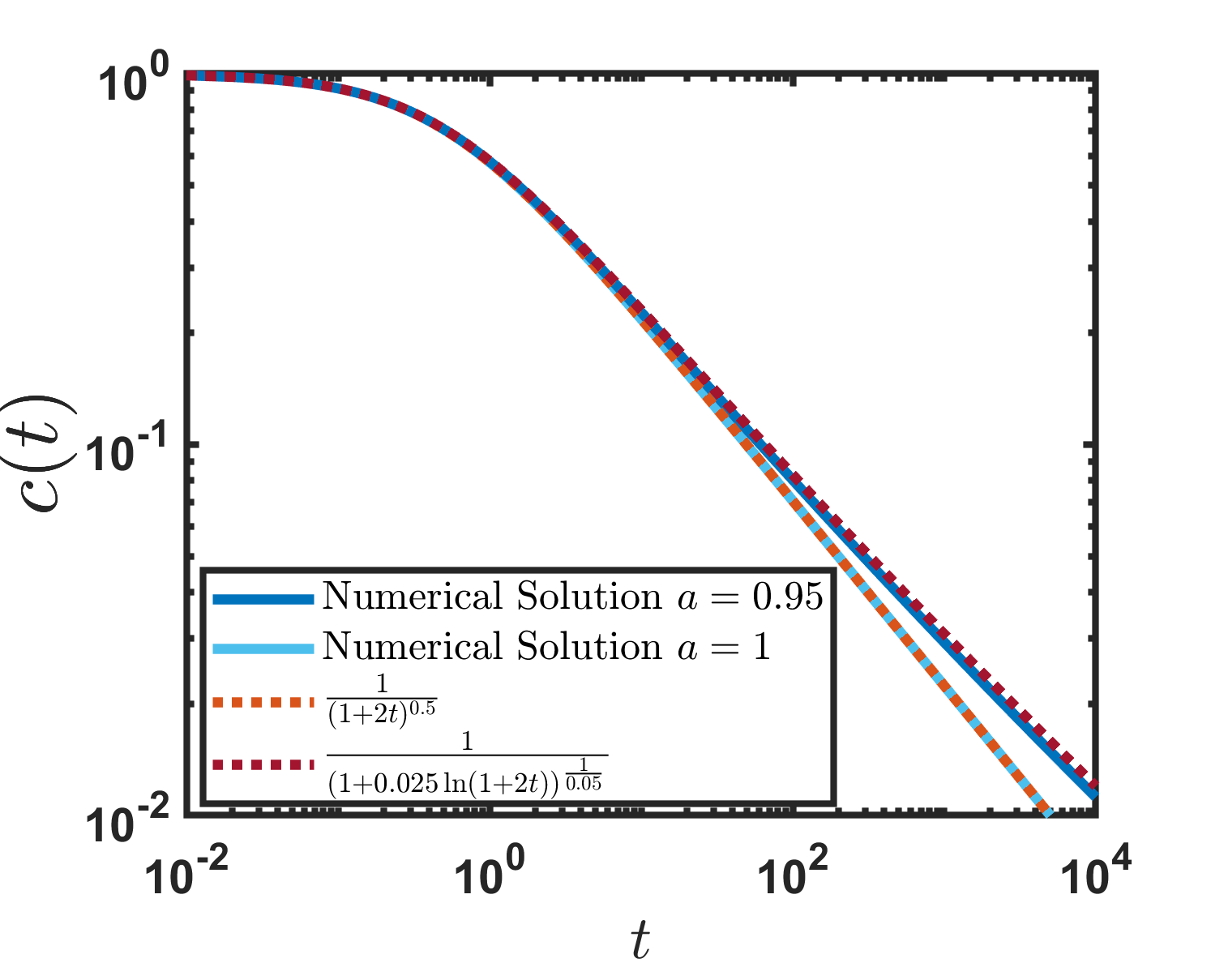}\q
\caption{Evolution of the total cluster density $c(t)$ for the SCS for $\lambda=1$. The numerical solution of Eqs.~\textbf{[17]}  for $\lambda=1$ is compared with the theoretical estimate \eqref{Ntotal} for $a=0.95$ (left panel) and $a=0.5$ (right panel). To plot the asymptotic relation \eqref{Ntotal} for $t\sim {\cal O} (1)$ we use the fitting
constant. For comparison the evolution of $c(t)$ for $\lambda=1$ and $a=1$ is also shown along with the theoretical
prediction, Eq.~\eqref{eq:lam1}. }
 \label{Fig:Supll_7}
\end{figure}

\begin{equation}
\label{cktotal} c_k(t) \sim
\begin{cases}
t^{-1}                                   &a_2 <a \leq a_1 =1\\
t^{-1} \ln t                            & a=a_2 \\
t^{-\alpha_p }                       &a_{p+1} <a < a_p; \qquad k>p>2 \\
t^{-\alpha_p} \ln  t                &a = a_p; \qquad \qquad \quad k>p>2.
\end{cases}
\end{equation}
with $\alpha_p=(p^{a_p}+p^{a_p-1})/2$ and
\begin{equation}
\label{Ntotal} c(t) \sim
\begin{cases}
t^{-1/2 }                               &a = a_1 =1\\
\left(\ln t \right)^{-1/(1-a)}                &a_{p+1} <a < a_p \\
\left[\ln (t/\ln t) \right]^{-1/(1-a)}                    &a = a_p.
\end{cases}
\end{equation}

Thus we conclude that the time dependence of the densities $c_k(t)$  undergoes at $a=a_p$ discontinuous (first order)
phase transitions. At the same time the total density $c(t)$ demonstrates at the transition  points $a_p$ (except for
$a_1=1$) only logarithmically weak alterations of its time dependence.  Figure \ref{Fig:Supll_7} illustrates the
dependence $c(t)$ for $\lambda=1$ and $s=a-1$ for different $a$. The theoretical estimates \eqref{Ntotal} agree with
the simulation data.

We wish to stress that the above analysis of the weak first-order phase transitions refers to the systems in the
thermodynamic limit. In finite systems not only the abrupt change of the relaxation behavior of $c_k(t)$ would be
observed at $a=a_p$, but also an abrupt change of  the exponent $\delta$, characterizing the dependence of the final
number of clusters on the system size, $C \sim N^{\delta} $.

\section{Systems with partial disintegration}

We analyzed several models of partial  disintegration where an abundant amount of monomers are produced, e.g., we
explored a model where a significant part of an aggregate (say a half) is preserved, while the other part crumbles into
monomers. Here we sketch our analysis of a more symmetric random model defined as follows: A cluster of size $k$ either
breaks into $k$ monomers, or a dimer and $k-2$ monomers, or a trimer and $k-3$ monomers, etc., and all these events
occur with equal probabilities. The governing rate equations for this model read
\begin{subequations}
\label{eq:erosion}
\begin{align}
\dot c_{1} &= -A_1c_{1} - \sum_{j=1}^{\infty} A_j c_{j} + \sum_{j=2}^{\infty}
\left(\tfrac{j}{j-1}+\tfrac{j-2}{2}\right) R_{j} c_{j}
 \\
 \dot c_{k} &= A_{k-1}c_{k-1} - A_{k}c_{k} + \sum_{j=k+1}^{\infty} \frac{R_{j} c_{j}}{j-1}  -R_k
c_k . \label{cktauer}
\end{align}
\end{subequations}
Similarly to the case of complete disintegration, it is natural to exploit the homogeneous kernels $A_k= k^a$ and
$R_k=\lambda k^{r}$. For the exponents $a=0$ and $r=-1$, which corresponds to the previously studied case of $a=0$ and
 $s=-1$, the analysis similar to that for complete disintegration shows that there are two critical values,
$\lambda_{\rm low}= 2$ and  $\lambda_{\rm up}= 3$. Thus the critical interval is shifted towards larger $\lambda$. For
$\lambda<\lambda_{\rm low}= 2$, the system falls into a jammed state  which depends on initial conditions; for
$\lambda>\lambda_{\rm up}= 3$, an equilibrium state is reached. For the critical interval $2 \leq \lambda \leq 3$,  the
SCS is observed. The final densities are (see also Fig.~\ref{Fig:Erosion})
\begin{equation}
\label{eq:Ck_erosion} C_k(\lambda) =
\begin{cases}
c_k(\tau_\text{max})(1-\delta_{k,1})  & \lambda < 2 \\
0                                            & 2 \leq \lambda\leq 3 \\
\frac{\Gamma(k+1) \Gamma(1+\lambda) }{\Gamma(k+\lambda)} \frac{(\lambda-2)( \lambda-3)}{\lambda(\lambda-1)}  & \lambda
> 3.
\end{cases}
\end{equation}

\begin{figure}
\centering
\includegraphics[width=9.5cm]{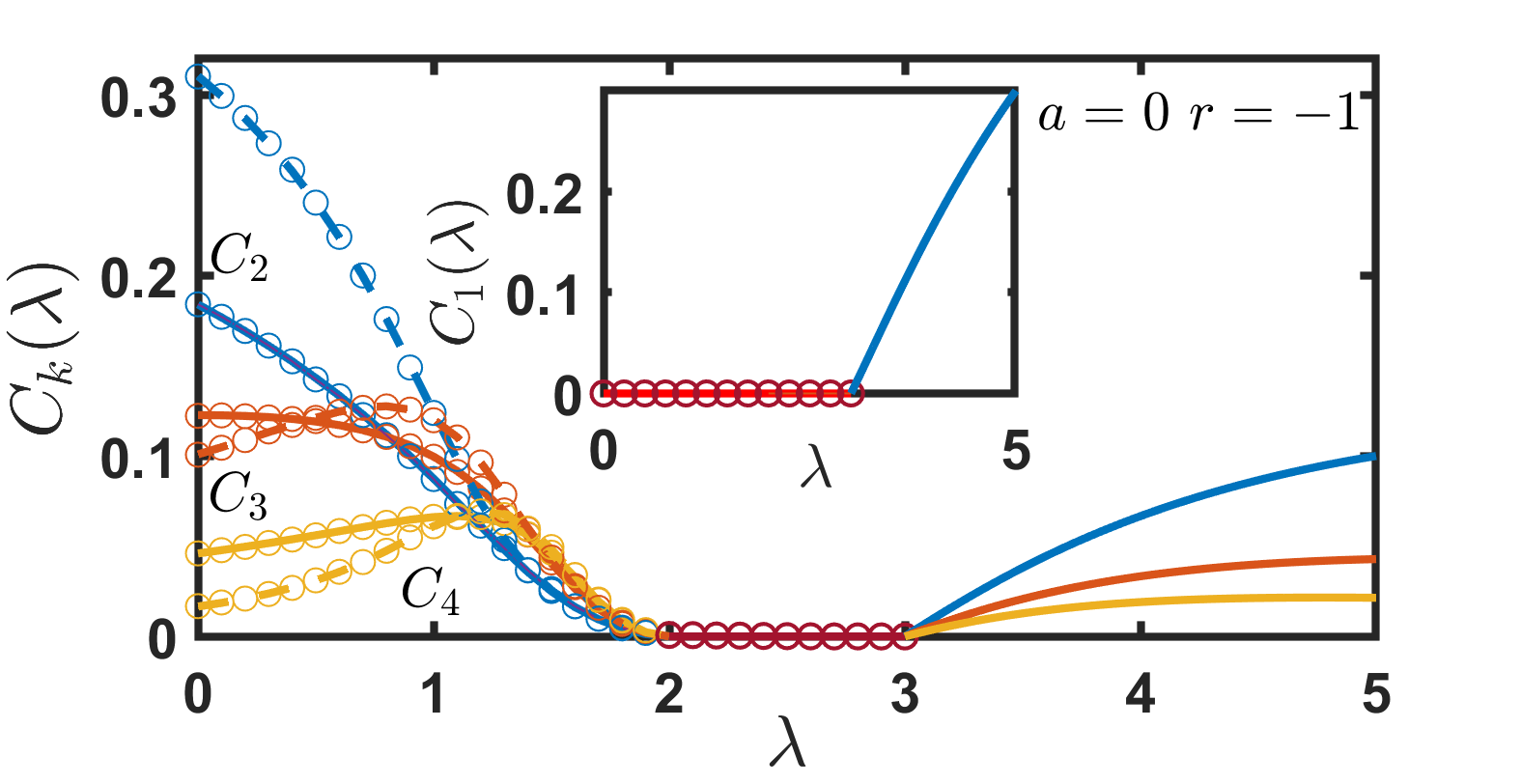}
\caption{The final densities for the model with partial disintegration, Eqs.~\eqref{eq:erosion} with rates $A_k= k$ and
$R_k=\lambda/k$. At $\lambda_{\rm low}=2$, this system undergoes a continuous phase transition from a jammed state to
the SCS; at $\lambda_{\rm up}=3$, it undergoes a continuous phase transition from the SCS to an equilibrium state. The
final densities in  a jammed state depend on initial conditions. Solid lines: $c_{1}(0) =1$; dashed lines: $c_{1}(0) =
0.2$, $c_{2}(0) = 0.4$.  Curves are solutions of rate equations; Monte Carlo results are shown by dots. Inset: The
density of monomers.}
 \label{Fig:Erosion}
\end{figure}

Similar results emerge for other values of the exponents $(a, r)$ and other breakage models; we investigated some
of these models analytically and numerically. Hence the SCS is generic for aggregating systems with disintegration.
Furthermore, the continuous transitions from the SCS to equilibrium state at the upper critical point $\lambda=
\lambda_{\rm up}$ and to jammed state at the low critical point $\lambda=\lambda_{\rm low}$ are also generic. Other
properties revealed for systems with a complete disintegration are also found for the case of partial disintegration.

\section{Monte Carlo simulations}

For the numerical analysis of finite systems we use a Monte Carlo method also known as Gillespie algorithm. Since the transition probability from one state to another depends exclusively on the present state, the reacting system can be presented by a Markov process. Each state $\{{\cal C}_1,{\cal C}_2, \ldots\} $ is
characterized by the number of aggregates of all sizes and the system can reach any of the following states
\begin{equation}
\label{eq:MCGill}
\begin{cases}
{\cal C}_1-1, \, {\cal C}_2, \ldots {\cal C}_k-1, \, {\cal C}_{k+1}+1, \ldots  & {\rm rate} \, \, \,  k^a     \\
{\cal C}_1+k, \, {\cal C}_2, \ldots {\cal C}_k-1, \, {\cal C}_{k+1},  \ldots  & {\rm  rate }\, \, \, \lambda k^s,
\end{cases}
\end{equation}
in the next step. The time of the next transition is chosen from a Poisson distribution with inverse average time
equals the sum of all reaction rates. The probability of a particular reaction from the set \eqref{eq:MCGill} equals to
the ratio of its rate and the sum of all rates. We simulated systems with up to $N = 10^8$ initial monomers.

\end{document}